\documentclass[english]{iopart}

\usepackage{graphicx}
\usepackage{url}
\usepackage{iopams} 

\expandafter\let\csname equation*\endcsname\relax
\expandafter\let\csname endequation*\endcsname\relax

\usepackage{amssymb}
\usepackage[latin1]{inputenc}
\usepackage{amsmath}
\usepackage{epsfig}
\usepackage{mathtools}
\usepackage{color}
\definecolor{darkblue}{rgb}{0, 0, .4}

\newcommand{\chiH}{\tilde{\chi}_{ _{ \scriptstyle H}}}
\newcommand{\chiL}{\tilde{\chi}_{ _{ \scriptstyle L}}}

\newcounter{todocounter}

\newcounter{fig}
\begin{document}

\title[Is the susceptibility of the Ising model differentially algebraic?]
{\Large Is the full susceptibility of the square-lattice Ising model 
a differentially algebraic function? }
\vskip 0.3cm


\author{A. J. Guttmann$^\ddag$, I. Jensen$^\ddag$,
J-M. Maillard$^\dag$, J. Pantone$^\pounds$}
\address{$^\ddag$ School of Mathematics and Statistics,
The University of Melbourne, Victoria 3010, Australia
}
\address{$^\dag$ LPTMC, UMR 7600 CNRS, 
Universit\'e de Paris 6,  
Tour 23, 5\`eme \'etage, case 121, 
 4 Place Jussieu, 75252 Paris Cedex 05, France} 
\address{$^\pounds$ Department of Mathematics, Dartmouth College,
Hanover, NH 03755, USA}

\vskip .3cm 

\vskip .3cm 

E-mail:  guttmann@unimelb.edu.au, ij@unimelb.edu.au, \\
$\qquad$ $\qquad$  maillard@lptmc.jussieu.fr, jay.pantone@gmail.com

\vskip .1cm 

\begin{abstract}

We study the class of non-holonomic power series  with integer 
coefficients that  reduce, modulo primes, or powers of primes, 
to algebraic functions. In particular we try to determine 
whether the susceptibility of the square-lattice Ising model 
belongs to this class, and more broadly whether the  susceptibility 
is a solution of a differentially algebraic  equation.

Initial results on Tutte's non-linear ordinary differential 
equation (ODE) and other simple quadratic non-linear ODEs
suggest that a large set of  differentially algebraic
power series solutions with integer coefficients  might reduce 
to algebraic functions modulo primes, 
or powers of primes. Since diagonals of 
rational functions are well-known to reduce, modulo primes, 
or powers of primes, to algebraic functions,
a large subset of differentially algebraic power 
series with integer coefficients 
 may be viewed as a natural ``non-linear'' 
generalisation of diagonals of rational functions.

Here we give several examples of
series with integer coefficients and non-zero radius 
of convergence that reduce to algebraic functions 
modulo (almost) every prime (or power of a prime). These 
examples satisfy differentially algebraic equations
with the encoding polynomial occasionally
possessing quite high degree (and thus
difficult to identify even with long series). These examples 
shed important light on the very nature of such differentially 
algebraic series.

Additionally, we have extended both the high- and 
low-temperature  Ising square-lattice susceptibility 
series  to 5043 coefficients. We find that even this long
series is insufficient to determine whether it reduces to 
algebraic functions modulo $3$, $5$, etc. This negative result 
is in contrast to the comparatively easy confirmation that the 
corresponding series reduce to algebraic functions 
modulo powers of $2$.

Finally we show that even with 5043 terms we are unable to 
identify an underlying differentially algebraic equation 
for the susceptibility, ruling out a number of possible
differentially algebraic forms. 

\end{abstract}


\vskip .1cm 

\noindent {\bf AMS Classification scheme numbers}: 03D05, 11Yxx, 
33Cxx,  34Lxx, 34Mxx, 34M55,  39-04, 68Q70   

\vskip .2cm


\vskip .2cm 
\vskip .1cm 

\section{Introduction}
\label{introduction}

The two-dimensional Ising model is arguably the most 
important statistical mechanical model ever conceived, 
historically, pedagogically, and in terms of its 
applications. Historically, the zeroth field-derivative 
of the free-energy was found by Onsager~\cite{O44} in 1944. The internal 
energy, given by the temperature derivative of the free-energy, can be 
expressed in terms of integrals of complete elliptic integrals of the 
third kind, or $\, _4F_3$ hypergeometric functions~\cite{Vish}, and is 
therefore  a {\em D-finite} or {\em holonomic} function. See 
\ref{appendix:defs} for a brief introduction to D-finite and 
D-algebraic functions, as well as some fundamental results about 
their behaviour.

The first field-derivative is the spontaneous magnetisation, and 
was publicly presented without proof by Onsager in 1949, and 
proved by Yang~\cite{Y52} in 1952. It is a simple algebraic function. 

The second field-derivative is the susceptibility $ \chi$, and no closed 
form expression is known. Despite this, we have a huge body of knowledge 
about the susceptibility. In 1976 Wu, McCoy, Tracy and 
Barouch~\cite{wu-mc-tr-ba-76} 
showed that it could be expressed as an infinite sum  of $ n$-dimensional 
integrals, $ \chi^{(n)},$ so that 
\begin{equation}
	\label{infinitesum}
	\chi  = \sum_{\mathclap{n \text{ even}}}\, \chi^{(n)}, \;\;    \, \,  
	\text{ for } T < T_c, \quad \text{ and } \quad     
	\chi  = \sum_{\mathclap{n \text{ odd }}}\, \chi^{(n)}, \;\; \, \, 
	\text{ for } T > T_c.
\end{equation}

 For the low-temperature susceptibility, only even powers of $ n$ 
contribute, starting at $n = 2$, while for the high-temperature 
susceptibility, only odd powers contribute. In 2001,  Orrick et al.~\cite{ONGP01} 
proved that each of these integrals $\chi^{(n)}$ is D-finite, but 
argued that their infinite sum, that is, the full susceptibility $\chi$ itself, 
is not. In fact, this follows from earlier work of Nickel \cite{Nickel99} who 
showed that the susceptibility has a {\em natural boundary} in the 
complex plane, and such functions cannot be D-finite. 
 
 In a series of papers, Maillard and 
co-workers~\cite{ze-bo-ha-ma-04, ze-bo-ha-ma-05c, bo-gu-ha-je-ma-ni-ze-08, bernie2010,High,Khi6} 
 found the linear ordinary differential equations (ODEs) satisfied by 
$\chi^{(3)}, \cdots , \chi^{(6)}$, and showed how, with sufficient 
computer power, others could, in principle, be found. $\chi^{(1)}$ is 
algebraic, and $\chi^{(2)}$ satisfies a low-order ODE,  but the 
order of the linear ODEs satisfied by $\chi^{(n)}$ increases 
rapidly with $n$ as does the degree of the polynomial coefficients 
of each derivative, so the number of series coefficients needed to 
discover the linear ODE also increases rapidly with $n$. In 2011, 
Chan et al.~\cite{CGNP11} 
extended the work of Orrick et al. to other two-dimensional 
lattices, and gave an expansion of the scaled form 
of the susceptibility to unprecedented accuracy.
  
Accepting that the susceptibility is not 
D-finite\footnote{No rigorous proof of this 
result exists, but no reasonable person doubts it.}, the question 
arises as to which class of functions the susceptibility 
belongs? In earlier work~\cite{Auto}, based on a series 
expansion of 2000 terms we showed that modulo powers of 2, one 
cannot distinguish the full susceptibility from some 
simple diagonals of rational 
functions\footnote{These are known to be D-finite -- see \ref{appendix:defs}.} 
that reduce to algebraic functions modulo $2^r$. 
However, for other primes we were unable to find 
corresponding algebraic function reductions. 

From another perspective, one may ask which class of functions  
is the natural generalisation of D-finite functions, and then 
try to determine whether or not the Ising susceptibility 
belongs to that class. The answer, or at least one answer, is 
the class of {\em differentially algebraic} 
functions.
 A series $\, F(x)$ is called 
a {\em differentially algebraic}, or simply {\em D-algebraic}, if there
 exists a polynomial  $P$ such that $F(x)$ satisfies a polynomial 
differential equation of finite order $k$: 
\begin{equation}
\label{1dalg}
P(x,\, F(x), \, F'(x), \,  \ldots, \,  F^{(k)}(x)) \,\, =\,\,\, 0.
\end{equation}
where $F^{(k)}(x)$ denotes the $k$-th derivative. Note that a $D$-finite 
function can also be seen as the solution of differentially 
algebraic ODEs. A relevant example is the ODE for the diagonal correlations 
$C(N,N)$ of the square Ising model. While this satisfies a $D$-finite ODE, 
it is also known to satisfy a differentially algebraic equation, 
notably\footnote[1]{$D$-finite functions can be expressed as solutions of (an 
infinite number of) differentially algebraic equations with movable singularities. 
However, that $D$-finite functions can also be solutions of differentially algebraic 
equations with {\em fixed critical points} is quite remarkable.} 
(a sigma-form of) Painlev\'e VI (see, for instance, equation (9) in~\cite{PainleveFuchs}).

In this paper we investigate the possibility that the susceptibility is a
differentially algebraic function. We tackle this from several 
perspectives. Firstly, we report on a dramatic extension 
of the susceptibility series from 
2000 terms to 5043 terms. We use this extended data in two different 
ways. The first is to provide further confirmation of the results 
found earlier~\cite{Auto}, that the susceptibility reduces to 
algebraic functions modulo $2^r,$ and unfortunately to discover that 
even with these much longer series we were unable to find a similar 
algebraic function reduction modulo other primes. 
If the susceptibility were holonomic, or indeed differentially algebraic, then 
with a sufficient number of terms we could, in principle,
 find the underlying ODE (given appropriate levels of computer power 
and precision). Then we could predict the next 100, or 1000 or whatever 
coefficients. If these agreed with the exact coefficients, then 
there would be no doubt that the exact ODE had been found. While 
that would not be a rigorous proof, it
  would undoubtedly be correct, and form the basis from which a 
rigorous proof could be constructed.
 
 The second approach is to use an algorithm, written 
by one of us~\cite{JP16}, to search for differentially 
algebraic functions. With such a long series, we are 
able to rule out algebraic differential equations of quite 
high order. Finding the susceptibility in this way would be 
almost too good to be true, so 
while disappointing, this negative result is not totally unexpected.
 
 In order to further explore the possibility that the susceptibility 
might be differentially algebraic, we have embarked on a study of 
the behaviour of such functions when their series expansions have 
{\em integer coefficients}, which is of course the case for the 
susceptibility. We study several examples of 
differentially algebraic functions whose series expansions have 
integer coefficients. For all the examples we choose, we find 
that the  series reduce modulo $p^r$ to an algebraic function 
(henceforth, we use ``modulo $p^r$'' to 
mean ``modulo a prime or a power of a prime''). There are known 
counterexamples to this observation in the form of series
with Hadamard gaps~\cite{Hadamard1,Hadamard2,Hadamard3}, 
such as the theta function $\theta_3(0,q)$. However, if their 
expansion variable is changed from the nome to the modulus, they 
are no longer counterexamples. This kind of lacunary series 
exception is not expected to be the kind of standard series 
we will encounter in lattice statistical mechanics.
 
 We also wish to make a connection between our earlier observation 
that the susceptibility reduces to algebraic functions modulo $ 2^r$ 
and the behaviour we observe in the examples of differentially algebraic 
functions we study. We first note that functions expressible as diagonals 
of rational functions~\cite{Short}, see (\ref{G4}), must 
be\footnote{Conversely, there is a conjecture due to Christol~\cite{Short} 
that a D-finite power series with integer coefficients and non-zero radius 
of convergence {\em is} the diagonal of a rational power series.} 
D-finite, as proved by Lipshitz~\cite{L88}. Secondly, as proved by 
Furstenberg~\cite{F87}, the diagonal of a rational power series 
must be an algebraic series modulo $p$ 
for almost all primes $p$.
 
 Our studies of miscellaneous examples of power series with 
integer coefficients, non-zero radius 
of convergence and no Hadamard gaps~\cite{Hadamard1,Hadamard2,Hadamard3},
 that are {\em differentially algebraic} show that they reduce, 
modulo most $ \, p^r$ to an algebraic function. This 
suggests that a large set of such series
might be seen as a natural generalisation of diagonals of 
rational functions. We provide abundant evidence that this is 
a reasonable, and useful, working hypothesis. 
 
 Connecting our examples to the susceptibility problem, we observe that 
for many simple examples it often requires thousands, or even tens 
of thousands, of coefficients to find the corresponding algebraic 
equations modulo even a small prime. This is consistent with the 
behaviour we have found for the susceptibility, and leaves open 
the question whether the susceptibility is differentially algebraic.

A broader aim of this paper is to understand the class of 
{\em non-holonomic} power series 
with integer coefficients that 
reduce, modulo (almost every) $p^r$ to algebraic functions~\cite{Gilles}.
 We showed~\cite{Selected} that a 
non-linear ODE, obtained by 
Tutte~\cite{Tutte5,Tutte6} for the generating function of 
the $ q$-coloured rooted triangulations 
by vertices,  belongs to this class, at least 
for $ q=4$. 

Not all   differentially algebraic power series
with  integer coefficients reduce to 
algebraic functions modulo almost every $p^r$.
For instance, 
the generating function of the  squares $ \sum_{n \ge 1} x^{n^2}$ 
(which amounts to considering the theta function as a function of 
the nome $q$, namely 
$\theta_3(0, q)  =  1  + 2  \sum_{n \ge 1} q^{n^2}$) 
is well-known to be differentially algebraic~\cite{Jacobi,Chakravarty} 
but does not reduce modulo $p^r$ to an algebraic 
function\footnote{The results are different if one 
views these functions {\em as functions
of the modulus $k$ instead of the nome $q$}.
For instance, the Eisenstein series $E_4$, when seen as
a function of the modulus $k$, is holonomic, and can be expressed 
in terms of complete elliptic integrals 
of the first kind $K(k)$, and thus reduces to algebraic functions
modulo primes and power of primes: 
$E_4(q) =  K(k)^4 \cdot (1-k^2  (k')^2)$.}, as 
a consequence of Cobham's~\cite{Cobham} 
theorem, which, in essence, states that  if a series has 
only coefficients $0$ or $+1$ it can be algebraic modulo 
two successive primes 
only if it is rational. We will, however, present a large 
accumulation of examples of differentially algebraic power 
series with {\em integer coefficients} 
({\em including divergent series}) 
which reduce modulo primes to algebraic functions.

The results found for Tutte's generating function~\cite{Selected}, 
with $q=4$, as well as the  accumulated examples referred to 
above, naturally raise the question whether, or to what extent, 
``{\em most}'' { differentially algebraic power series}
with {\em integer coefficients} reduce modulo almost 
every $p^r$ to algebraic functions. This would then suggest 
that ``most''  differentially algebraic power series with 
{\em integer coefficients} can be seen as some sort of 
{\em ``non-linear'' generalisation of diagonals
of rational functions}. Recall that diagonals of 
rational functions are holonomic. They are well-known 
to reduce to algebraic functions modulo $p^r$~\cite{Short}. 
Note that differentially algebraic power series with 
 integer coefficients are an obvious generalisation of
D-finite  series with integer coefficients, and such D-finite 
series appear to be algebraic modulo almost all $p^r$ 
(see Conjecture 12 and Section 5 in~\cite{European}).
The difference between diagonals of rational 
functions and D-finite  series with integer coefficients, 
corresponds to ``Christol's conjecture'', see~\cite{Short}, 
and for instance the ${}_3F_2$ hypergeometric example in 
Section \ref{noreduction}. 

The class of differentially 
algebraic power series has a very important property: 
it is {\em closed under the composition of 
functions}\footnote{The {\em composition of two differentially 
algebraic functions is differentially algebraic}, see Remark 4.3 
in Lipshitz and Rubel~\cite{Lipshitz} and, in the particular 
case of constructible differentially algebraic series,
 Bergeron and Reutenauer~\cite{Bergeron}. We thank D. Bertrand for 
providing a proof of Ehud Hrushovski of that result, in the most 
general framework. Note that differentially algebraic functions are 
{\em not closed under Hadamard product, Laplace transform, or
inverse Laplace transform}~\cite{Lipshitz}.
}. The class of non-holonomic power series with integer 
coefficients that reduce to algebraic functions 
modulo $p^r$ is probably also closed under the composition 
of functions.

A subset, the class of differentially algebraic power series with 
integer coefficients  that reduce to algebraic functions modulo $p^r$ 
is an extremely large class.
We will see in this paper that 5043 coefficients for 
the susceptibility series of the square-lattice
Ising model is still  insufficient to see if the square-lattice Ising 
susceptibility series reduces to an algebraic function modulo 
$p^r$ for $p \neq 2$. This suggests that trying to see if 
the Ising susceptibility series is a differentially algebraic power 
series might be a simpler question to address.  

\vskip .1cm 

\section{Revisiting Tutte's differentially algebraic generating function.}
\label{Tutte}

Before trying to find new results on the square-lattice Ising 
susceptibility, we will revisit~\cite{Selected} a simpler 
(but already important and non-trivial) 
example of a {\em non-holonomic} function
namely Tutte's generating function for $ q$-coloured rooted 
triangulations by vertices~\cite{Tutte5,Tutte6},
\begin{eqnarray}
\label{firstterms}
\hspace{-0.95in}&&  \quad \quad \,  \, \, \,
H(x) \, \, \, = \, \, \, \,  \, 
q \cdot \, ( q-1 ) \cdot \,  {x}^{2} \, \, \,  \, 
+ \, \, q \cdot \, ( q-1 )  \cdot \,  ( q-2 ) \cdot \, 
\sum_{n=3}^{\infty} \, \, P_n(q) \cdot \,  x^n, 
\end{eqnarray}
considered from the perspective of its reducability
 modulo $p^r$.  The coefficients of the series (\ref{firstterms})
can be obtained from a simple 
{\em quadratic recurrence relation}:
\begin{eqnarray}
\label{Tutterecq1}
\hspace{-0.95in}&&   \quad  \, \, \,\,
q \cdot \, (n+1)\cdot \,(n+2) \cdot \, h_{n+2} \, \,\, \, = \, \,  \, \,\,\,
q  \cdot \, (q\,-4)  \cdot \, (3\, n\, -1) \cdot \, (3\, n\, -2) \cdot \, h_{n+1} 
\nonumber  \\
\hspace{-0.95in}&&   \quad  \quad \quad \quad  \quad   \quad \quad  
\, + \, 2 \, \sum_{i=1}^{n} \, \,  i \cdot \, (i+1) \cdot \, 
(3\, n\, -3\, i\, +1) \cdot \,  h_{i+1} \, h_{n-i+2}, 
\end{eqnarray}
that can also be rewritten as 
\begin{eqnarray}
\label{Tutterecq2}
\hspace{-0.95in}&&   \quad  \, \, \,\,
q  \cdot \, (q\,-4)  \cdot \, \Bigl( 9 \, \,  n\, (n+1)\cdot \,  h_{n+1} 
\, \, -18  \, \,  (n+1) \cdot \,  h_{n+1} \,\,  +20 \cdot \,  h_{n+1}  \Bigr) 
\nonumber  \\
\hspace{-0.95in}&&   \quad   \quad  \quad  \quad 
\, + \, 2 \, \sum_{i=1}^{n} \, \,  \Bigl( i \cdot \, (i+1) \cdot \,h_{i+1}  \Bigr)  
\cdot \,  
\Bigl( 6 \cdot \, (n\, -i\, +2) \cdot \,  h_{n-i+2} \, - 10 \cdot \,  h_{n-i+2} \Bigr)
\nonumber  \\
\hspace{-0.95in}&&   \quad   \quad \quad \quad \quad  \quad   \quad \quad
\,\, - \, q \cdot \, (n+1)\cdot \,(n+2) \cdot \, h_{n+2} 
\, \,\, \, = \, \,  \, \,\,\, 0. 
\end{eqnarray}
The generating function (\ref{firstterms})
is a solution of the 
{\em differentially algebraic equation}~\cite{Selected}: 
\begin{eqnarray}
\label{Tutteq}
\hspace{-0.95in}&&   \quad  \, 
q \cdot \, (q\, -4)  \cdot \, 
\Bigl(9 \, x^2 \, { {d^2 H(x)} \over {dx^2}} \,
 - \,18 \, x \, { {d H(x)} \over {dx}} \,
 +  20 \, H(x) \Bigr)
 \nonumber  \\
\hspace{-0.95in}&&   \quad  \quad    \quad  
+ \, \Bigl( 6 \, x \, { {d H(x)} \over {dx}}
\, - 10 \, H(x) \, -q\, x  \Bigr)
 \cdot \,   { {d^2 H(x)} \over {dx}^2} 
\, \,\,  =  \, \, \, \,  \, 2 \, q^2 \cdot \, (1\, -q)  \cdot \, x. 
\end{eqnarray}
It is quite easy to deduce
(\ref{Tutteq}) from (\ref{Tutterecq2}), the rhs of (\ref{Tutteq}) 
depending on the initial conditions, namely 
the first terms of the series (\ref{firstterms}).

\vskip .1cm

In~\cite{Selected} we have shown that, in the $ q  =  4$ case,
this generating function 
\begin{eqnarray}
\label{serq4}
\hspace{-0.95in}&&  \, \, \quad     
H(x) \, \, = \, \, \,  \,  \, 
12\,{x}^{2}  \, \, +24\,{x}^{3} \, +168\,{x}^{4} \, +1656\,{x}^{5} 
\, +19296\,{x}^{6} \, +248832\,{x}^{7} \, 
\nonumber \\
\hspace{-0.95in}&& \qquad \quad \quad \quad 
 +3437424\,{x}^{8} \, 
\, +49923288\,{x}^{9} \, +753269856\,{x}^{10} 
\,\, \, \, + \, \, \cdots 
\end{eqnarray} 
is {\em non-holonomic} and {\em reduces to an
algebraic function modulo every prime or power of a prime}.
Let us consider other solutions of (\ref{Tutteq}) for $ q=  4$,
corresponding to different initial conditions\footnote[1]{Note 
a missprint in~\cite{Selected}: one should 
read ``$ h_{0}   =  0$ 
but does not impose $ h_{1}   =  0$'' instead of 
``$ h_{0}   =  0$ and $ h_{1}   =  0$, but 
does not impose $ h_{2}   =  q  (q-1)  =  12$''.}
namely the one-parameter family of solutions 
(here $ A$ denotes that parameter):
\begin{eqnarray}
\label{serq4onepara}
\hspace{-0.95in}&&  \, \, \,    \quad  
H_A(x) \, \, \, = \, \, \, \,\, 
(A-1) \cdot \, x \,\,   \, +12\, {{x^{2}} \over {A}}\,\,   \, 
+24\, {{x^{3}}  \over {A^3}} \, \, \,  
+168\, {{x^{4}}  \over {A^5}}\, \,  \, 
+1656\, {{x^{5}}  \over {A^7}}\, \,\,   + \,\,   \cdots 
\end{eqnarray}
This one-parameter family of solutions can be written 
\begin{eqnarray}
\label{serq4onepar}
\hspace{-0.95in}&&  \, \, \quad  \quad  \quad     \quad  
H_A(x) \, \, \, = \, \, \, \,\, \, 
 -x \, \, \, \, \, 
+ \, A^3 \cdot \Bigl( {{x} \over {A^2}} 
\, \, + \,   H \Bigl( {{x} \over {A^2}} \Bigr) \Bigr), 
\end{eqnarray}
where the function $ H(x)$ in (\ref{serq4onepar})
 is the series (\ref{serq4}).
Since (\ref{serq4}) reduces to algebraic functions modulo 
each prime or power of a prime, 
one sees immediately that these other solutions 
(\ref{serq4onepara}) of the non-linear differential 
equation (\ref{Tutteq}) for $ q=4$, 
{\em also reduce to algebraic functions modulo every prime or power of a prime}.

\vskip .1cm

\subsection{Tutte's differentially algebraic generating function 
for other values of $ q$.}
\label{Tutte1}

It is tempting to see if one also has reduction to algebraic 
functions modulo $p^r$ for values of $q$ other than $q=4$. Tutte's 
algebraic differential equation (\ref{Tutteq}) is 
known~\cite{Odlyzko,Bernardi} to 
provide algebraic solutions for all the Tutte-Beraha numbers, 
and in fact, for all $q$ of the 
form $q = 2 +2\cos(j\pi/m)$ ($j$ and $m$ are integers). Therefore 
we have considered Tutte's differentially 
algebraic equation (\ref{Tutteq}) for miscellaneous values
that {\em are not} of the form 
$q  = 2 +2  \cos(j\pi/m)$, for instance
integer values larger than $4$, $ q= 5, 7, ...$,
or rational values $ q  = M/N$.

For instance, for $ q  =   5$,
one finds that the series (\ref{firstterms}) still reduces 
to algebraic functions modulo every prime. Modulo $ 7$ the 
series (\ref{firstterms}) for $ q=  5$, namely 
$\,  6\,x^2 \, +4\, x^3$
$ \, +2\, x^4 \, +3\, x^5 \, 
+5\, x^6 \, +6\, x^7 \, +5\, x^8 \, +5\, x^9 \, + \, \, \cdots$, 
is a solution of the polynomial equation 
\begin{eqnarray}
\label{q5ser7}
\hspace{-0.95in}&&  \, \, \quad  \quad   
F(x)^{5} \,  \, + \, (5\,{x}^{2} +6\, x+6)  \cdot \, F(x)^{4} \, 
\, + \, ({x}^{4}+2\,{x}^{3}+2\,{x}^{2}+4\,x+4)  \cdot \, F(x)^{3} 
\nonumber \\ 
\hspace{-0.95in}&&  \, \, \quad   \quad   \quad   \, 
+{x}^{2} \cdot \, (6\,{x}^{2} +6\,x +1)  \cdot \, F(x)^{2}  \, 
\, +{x}^{2} \cdot \, (2\,{x}^{5} +5\,{x}^{3} +5\,{x}^{2} +x +5)  \cdot \, F(x)
\nonumber \\ 
\hspace{-0.95in}&&  \, \, \quad   \quad  \quad    
\,  +  {x}^{4} \cdot \, (2\,{x}^{5} +4 \,{x}^{4} 
+3 \,{x}^{3} +\,{x}^{2} +2\,{x} +5)
\, \, \,  = \, \, \, \,  0.
\end{eqnarray}
Modulo $ 11$ the $ q=  5$ series, namely 
$\, 9\,x^2 \, +5\, x^3 \, +x^5 \, +7\, x^6 
\, +9\, x^7 \, +10\, x^8 \, +8\, x^9 \, + \, \cdots$,
 is a solution of the polynomial equation 
\begin{eqnarray}
\label{q5ser11}
\hspace{-0.95in}&&  \, \, \quad 
F(x)^{3} \, 
\, +7 \cdot \, F(x)^{2} \, 
\, +x \cdot \, (4\,{x}^{2}+7\,x+9) \cdot \, F(x) \, \,  
+x^3 \cdot \, (2 \, +3 \, x \, +10 \, x^2)
\, \,  \, = \, \,  \, \, 0,
\nonumber
\end{eqnarray}
and modulo $ 13$ the $ q=5$ series 
is a solution of a polynomial equation of degree 
$ 15$ in $ F(x)$, and degree $27 $ in $ x$. 
Similarly to $ q=4$ (see (\ref{serq4onepara})),
the {\em other solutions} of (\ref{Tutteq}) for $ q=5$
also reduce to algebraic functions modulo every prime,
or power of a prime.

We give in \ref{appendix1} many other examples of reductions
modulo primes of the solutions of (\ref{Tutteq}) for 
various values of $ q$ that {\em are not} of the 
form $ q  =   2  +2  \cos(j \pi/m)$.

Tutte's differentially algebraic equation (\ref{Tutteq}) provides 
a non-trivial, important but simple enough, ``toy example'' of
reduction modulo primes or powers of primes, of
differentially algebraic functions. We also give in \ref{appendix2}
an example of a (quite natural) generalisation of Tutte's 
differentially algebraic equation (\ref{Tutteq}) that also yields 
power series solutions with integer coefficients reducing 
to algebraic functions modulo $p^r$.

We have accumulated similar results for large sets of solutions 
of miscellaneous differentially algebraic functions having
power series solutions with integer coefficients, in particular 
power series {\em generated by simple quadratic recursions} 
(similar to the one
generating the solution series (\ref{firstterms}),  
(see equation (5) in~\cite{Selected})). Such examples 
are displayed in~\ref{appendix3}. One sees from these examples
that, even if these differentially algebraic power series
are {\em divergent series}, they still {\em  reduce to 
algebraic functions modulo primes or powers of primes}. 

These exploratory results on miscellaneous non-linear equations
suggest that  {\em differentially algebraic
power series solutions with integer coefficients and no Hadamard gaps 
usually reduce to algebraic functions modulo primes, 
or powers of primes}. If that is the case,
recalling that diagonals of 
rational functions are well-known to reduce to algebraic 
functions modulo $p^r$~\cite{Short},
 then differentially algebraic power 
series {\em could be seen as a natural ``non-linear'' 
generalisation of diagonals of rational functions}.

\vskip .1cm

\section{Brain storming: square-lattice Ising susceptibility 
 and diagonals of rational functions.}
\label{susceptdiag}

One of the simplest
examples of a differentially algebraic equation is  the 
Chazy III equation. It is  worth considering as a toy example 
of some properties of the Ising susceptibility because 
{\em it has the Painlev\'e 
property}\footnote{There is little need to emphasise 
the importance of the Painlev\'e property to the 
square-lattice Ising model: the diagonal correlation functions
are known~\cite{PainleveFuchs,FuchsPainleve} to be solutions of 
(a $\, \sigma$ form of) Painlev\'e VI.}, 
and because its solutions can be expressed as 
{\em ratios of holonomic functions} (see \ref{G4}). Recall
 that the {\em ratio} (not the product!)
 of {\em two holonomic} functions is generically {\em non-holonomic}.
The ratio $\tau(x)  = {y_1}/{y_2}$
of two solutions of the linear ODE, 
${{d^2 y} \over {dx^2}} +R(x) \cdot  y  = 0$, 
is a solution of the Schwarzian equation
$ \{\tau(x), x\} = 2  R(x)$. Considering ratios of holonomic 
functions is thus a very simple way to build (selected) 
non-holonomic examples. In fact, these solutions can be expressed as 
a ratio of diagonals of rational functions, which implies that 
they have (circular) {\em natural boundaries}~\cite{Chazy1,Chazy2}, 
which is  also observed to be a property of the susceptibility 
of the Ising model~\cite{bo-gu-ha-je-ma-ni-ze-08}. 

\vskip .1cm

The {\em Chazy III equation}~\cite{Chazy1,Chazy2} is a 
third-order non-linear differential equation the solutions of 
which have a natural boundary~\cite{Chakravarty}. It is
\begin{eqnarray}
\label{Chazy}
\hspace{-0.85in}&& \quad \quad \quad \quad \quad \quad \quad 
{{d^3 y} \over {dx^3}}
 \,\, \, \,  = \, \, \, \,\,  
2 \, y {{d^2 y} \over {d x^2}} \, \, \, 
 \,   \,   - 3 \, \left({{d y} \over {d x}} \right)^2, 
\nonumber 
\end{eqnarray}
which can be rewritten in terms of a 
{\em Schwarzian derivative}~\cite{Schwarzian,Schwarzian2}
\begin{eqnarray}
\label{Chazy2}
\hspace{-0.95in}&& \quad \, \quad \quad \, 
 \, f^{(4)} \, \, = \, \,\, \,  2 \, f'^2 \cdot \, \{f, \, x\} 
\, \, = \, \, \, \, 
2 \, f' \, f'''\, - 3 \, f''^2   \, \, \quad \quad
\hbox{with} \quad \quad  \,
y \, = \, \, {{d f} \over {dx}},
\end{eqnarray}
where $ \{\cdot, \cdot \}$ denotes the Schwarzian derivative.

\vskip .1cm

In~\cite{Selected} it has been remarked that ratios of particular
holonomic functions, namely 
{\em ratios of diagonals of rational functions},
 automatically reduce to algebraic functions modulo
every prime, or power of a prime, which is also a property 
we observed for the susceptibility series modulo powers of $2$. 

\vskip .1cm

The {\em non-holonomic} susceptibility series is known 
to be an {\em infinite sum} of holonomic 
functions~\cite{bo-gu-ha-je-ma-ni-ze-08,wu-mc-tr-ba-76}, namely
the $n$-fold 
integrals~\cite{ze-bo-ha-ma-04,ze-bo-ha-ma-05c,bernie2010,High,Khi6,CalabiYauIsing} 
 $\chi^{(n)}$ that 
{\em are themselves diagonals of rational 
functions}~\cite{Short,Cobham,Poorten,Denef,Adamczewski}. 
To some extent, the remarkable result~\cite{Auto} that 
the non-holonomic susceptibility series reduces to 
algebraic functions modulo $2^r$ can be seen as 
a property of these diagonal of rational functions, 
namely that these  $\chi^{(n)}$ reduce to zero 
modulo $2^r$ when $n$ is large enough~\cite{Selected}. 

\vskip .1cm  

All these ideas suggest that it may be useful to take a fresh 
look at the square-lattice Ising susceptibility from the perspective 
of a ``{\em diagonals of rational function approach}''~\cite{Short}.
 The ``Chazy III scenario'' just discussed, suggesting
that the square-lattice Ising susceptibility might just be a 
{\em ratio} of diagonals of rational functions reducing to 
algebraic functions modulo primes, and, possibly the occurrence 
of natural boundaries, is certainly far too naive. If that 
were the case, then the Ising susceptibility
would easily be seen to reduce to algebraic functions 
{\em not only modulo} $ 2^r$, but also modulo any power 
of a prime, which does not seem to be the case~\cite{Auto}. 
Accordingly we explore  more involved scenarios.

\vskip .1cm

\section{More involved constructions from diagonals of rational functions.}
\label{moreinvolved}

If ratios of diagonals of rational functions
are automatically such that they reduce to algebraic functions 
modulo $p^r$ and are also differentially algebraic functions
(see Section 5.2 and Appendix D in~\cite{Selected}), one can easily 
imagine more involved expressions of diagonals of rational functions
that will also reduce  to algebraic functions modulo
$p^r$~\cite{Selected}. Note that the class of rational expressions 
of diagonals of rational functions reduces to the previous class 
of ratios of diagonals of rational functions. This is a 
consequence of the fact that polynomial expressions of diagonals 
of rational functions are themselves diagonals of rational 
functions. To get a larger class, one needs to consider, at least,
{\em algebraic functions} of diagonals of rational functions 
(as already suggested in~\cite{Selected}).  We look at some of 
these in the next subsection.

\subsection{Algebraic functions of diagonals of rational functions. \\}
\label{moreinvolvedalg}

Algebraic functions of diagonals of rational functions, which are
generically non-holonomic, reduce to algebraic functions 
modulo $p^r$ and are also differentially algebraic.

Let us just give a simple heuristic and pedagogical example. 
Consider two ${}_2F_1$ hypergeometric functions
which are also diagonals of rational functions
\begin{eqnarray}
\label{twodiag}
\hspace{-0.95in}&&   \,  \,
H_1 \, = \, \, _2F_1\Big(\Bigl[{{1} \over {2}}, \, {{1} \over {2}}\Bigr], \, [1], 
\, 16 \cdot 20 \cdot \, x^2\Bigr), \quad \, \, \,
H_2 \, = \, \, _2F_1\Bigl(\Bigl[{{1} \over {3}}, \, {{1} \over {3}}\Bigr], \, [1], 
\, 27 \cdot 20 \cdot \, x^2\Bigr),
\end{eqnarray}
and consider the two roots of 
\begin{eqnarray}
\label{roots}
\hspace{-0.95in}&&   \quad  \quad \quad 
\quad \quad \quad \quad \quad \quad 
 z^2 \,\,  - \,\, 2   \, H_1 \cdot \, z \,  \,\,+\,\, H_2  
\,  \,\, = \, \, \,0, 
\end{eqnarray}
namely 
$z_{\pm} = H_1 \pm  (H_1^2 -H_2)^{1/2}$,
which have series expansions 
{\em with integer coefficients}
\begin{eqnarray}
\label{rootsser}
\hspace{-0.95in}&& \quad    
z_{-}(x)=\,\, \,\, \, 
1\,\,\,  \,  -10\,x\,\,\,+80\,{x}^{2}\, -1040\,{x}^{3}\,
+14400\,{x}^{4}\, -145920 \,{x}^{5}\, +3200000\,{x}^{6}
\nonumber \\ 
\hspace{-0.95in}&&   \quad  \quad  \quad  
\, -10992320\,{x}^{7} \, +784000000\,{x}^{8} \, 
+6780473600\,{x}^{9}  \, + 203212800000\,{x}^{10}
\nonumber \\ 
\hspace{-0.95in}&&   \quad \quad \quad 
\, +5987941079040\,{x}^{11}
\,  +54641664000000\,{x}^{12} 
\,+3543158723957760\,{x}^{13} 
\nonumber \\ 
\hspace{-0.95in}&&   \quad \quad \quad 
\,+15076638720000000\,{x}^{14}
\,\,\,  \,+ \,\, \cdots
\end{eqnarray}
and $  z_+(x) =  z_-(-x)$. Apart from   
the coefficients of $x$, $x^3$, $x^5$, $x^7$, 
all the coefficients are positive integers.

These two series are {\em not holonomic} (square roots of 
holonomic functions are generically {\em not holonomic}), 
rather they  {\em are differentially algebraic}. 
The expression $H_1^2-H_2$ in (\ref{roots}) is 
holonomic (it is a simple polynomial of holonomic functions), 
being the solution of a linear differential 
operator $L_5$ of order five. The square root of that holonomic 
function $g(x),$ defined by $g(x)^2= H_1^2-H_2$, is  
differentially algebraic, and is the solution of the 
polynomial differential equation $L_5(g(x)^2) = 0$. 
Straightforward (but tedious) differential algebra 
calculations show that the series (\ref{rootsser}) is the 
solution of the order-seven polynomial differential equation
\begin{eqnarray}
\label{AH2mod5}
\hspace{-0.95in}&&   \quad    
P(x, \, F(x), \, F'(x), \, F''(x), \, F'''(x), \, 
F^{(4)}(x), \, F^{(5)}(x), \, F^{(6)}(x), \, F^{(7)}(x))
  \, \,  \,= \, \,  \,\, 0, 
\end{eqnarray}
where $F(x)$ denotes  the series  (\ref{rootsser}), and 
where  $F^{(n)}(x)$ denotes its $n^{th}$derivative. 
The polynomial $P$ is the sum of  $3769 $ monomials 
of degree at most $46$ in $x$,  $4$ in $F(x)$, 
  $F'(x)$, $F''(x)$,  and $F'''(x)$,  $3$ in $ F^{(4)}(x)$ and 
  $F^{(5)}(x)$,  $2$ in $F^{(6)}(x)$, and $1$ in $ F^{(7)}(x)$.
The complexity of this polynomial should be compared with the 
 simplicity of the original function  (\ref{roots}). 

 The polynomial equation (\ref{AH2mod5}) can
be written as:
\begin{eqnarray}
\label{AH2mod5bis}
\hspace{-0.95in}&&   \quad \quad \quad \quad \quad \quad 
\sum_{n=0}^{46}  \, \,   
x^n \cdot \, P_n(F(x), \,F'(x), \, \cdots, \,\, \ F^{(7)}(x))
 \, \, \, \ =\,  \,\, \ \, \, 0, 
\end{eqnarray}
where the polynomials $P_n$ are homogeneous polynomials of degree 
$11$ in $F(x), F'(x), \cdots F^{(7)}(x)$. If one
 scales $F(x)$ as $F(x) \rightarrow A \cdot F(x)$,
(and consequently $F^{(i)}(x) \rightarrow A \cdot  F^{(i)}(x)$), 
the polynomial $P$ in (\ref{AH2mod5})
scales as $P \rightarrow A^{4} \cdot P$. Therefore, if 
$F(x)$ is a solution of (\ref{AH2mod5}), 
$A \cdot F(x)$ is also a solution of (\ref{AH2mod5}). 

It is straightforward to see that the non-holonomic 
series (\ref{rootsser}) {\em  reduces to an
algebraic function modulo $p^r$} (see \ref{AFD}).

The class of algebraic functions of diagonals of rational 
functions is already a very large one. However, we will see 
in the next section (\ref{composidiag}) that  power series 
with integer coefficients reducing to algebraic functions 
modulo every prime, or power of a prime, corresponds to 
a  {\em much larger class}.

\vskip .1cm

\subsection{Compositions of diagonals of rational functions.\\ }
\label{composidiag}

\vskip .1cm

Let us consider another simple example corresponding to the 
{\em composition of two holonomic functions} (in fact,
the composition of two diagonals of rational functions).

Before introducing that example,  recall the 
{\em closure properties of  holonomic functions}. They 
are closed under the operations of addition, 
multiplication, indefinite integration, differentiation and 
{\em right composition with algebraic functions}. In other words 
if $H(x)$ denotes a holonomic function, and $A(x)$ an 
algebraic function, $H(A(x))$ is {\em necessarily holonomic}. 
In constrast, if one performs the composition of two holonomic 
(but non-algebraic) functions, the result is generically 
{\em not holonomic}. For instance, the reciprocal function 
$ R(x) = 1/x  $ is holonomic, but   the composition of $R(x)$
with a holonomic function $ H(x)$, namely
$ R(H(x))= 1/H(x)$  is generically  {\em non-holonomic}.
These closure properties 
{\em extend to diagonals of rational functions}. 
For instance, the class of diagonals of rational functions is 
also closed under the operations of addition, subtraction,
multiplication, indefinite integration, differentiation and 
{\em right composition with algebraic functions}. 

Consider the two holonomic functions (that are diagonals 
of rational functions~\cite{Short})
\begin{eqnarray}
\label{H1H2}
\hspace{-0.95in}&&   \quad  \quad  \, \, \,  \,  \, \,   
 H_1(x) \, = \, \,
 _2F_1\Bigl(\Bigl[{{1} \over {2}}, \, {{1} \over {2}}\Bigr],\, [1], 
\, 16   \, x  \Bigr) , 
\quad \quad  \, \, H_2(x) \, = \, \, x   \, H_1(x), 
\end{eqnarray}
and consider the following composition of these two 
holonomic functions:
\begin{eqnarray}
\label{compoH1H2}
\hspace{-0.95in}&&   \quad \quad   \quad  \, 
 H_1(H_2(x)) \,\, \,  = \, \,\, \, 
 _2F_1\Bigl(\Bigl[{{1} \over {2}}, \, {{1} \over {2}}\Bigr],\, [1], \, 
16  \, x \cdot  \, _2F_1\Bigl(\Bigl[{{1} \over {2}}, \, {{1} \over {2}}\Bigr],
\, [1], \, 16   \, x  \Bigr)\Bigr). 
\end{eqnarray}
This function has the following series expansion with 
{\em integer coefficients}:
\begin{eqnarray}
\label{seriescompoH1H2}
\hspace{-0.95in}&&   \quad  
 H_1(H_2(x)) \,\, = \,\, \, \,  1\,\,\, +4\,x\,\, 
+52\,{x}^{2}\,+832\,{x}^{3}\,
+14468\,{x}^{4}\,+263072\,{x}^{5} \,+4919728\,{x}^{6}
\nonumber \\ 
\hspace{-0.95in}&&   \quad  \quad  \quad  
\,+93824000\,{x}^{7}\,+1815689828\,{x}^{8}\,+35542852576\,{x}^{9}\,
+702276985968\,{x}^{10}
 \nonumber \\ 
\hspace{-0.95in}&&   \quad  \quad \quad  
 \,+13984093836288\,{x}^{11}\,+280299095853776\,{x}^{12}\,
+5650349273844992\,{x}^{13}\,
 \nonumber \\ 
\hspace{-0.95in}&&   \quad  \quad \quad  
 \, +114466793551793216\,{x}^{14}\,
+2329040212651647488\,{x}^{15} \, 
 \,\,\, \, +\, \, \cdots 
\end{eqnarray}
The coefficients of this series grow like $ \lambda^n$ 
where $ \lambda \simeq  21.7257468152791$.
The radius of convergence of this series $  1/\lambda$, 
is $ x_c \simeq  0.046028337184$, which is expected to be
a transcendental value\footnote{For a  transcendental condition 
like (\ref{seriescompoH1H2-xc}), one might expect, at first sight, 
an {\em infinite number} of (complex) transcendental values 
for $ x_c$. This is not the case here.} such that:
\begin{eqnarray}
\label{seriescompoH1H2-xc}
\hspace{-0.95in}&&   \quad  \quad   \quad   \quad  \quad  \quad 
16   \, x_c \cdot \, _2F_1\Bigl(\Bigl[{{1} \over {2}}, \, {{1} \over {2}}\Bigr],
\, [1], \, 16   \, x_c  \Bigr)
 \, \,\, =\, \, \, \, 1.  
\end{eqnarray}
It is reassuring that, despite the fact that the series 
(\ref{seriescompoH1H2}) is not holonomic, nevertheless a
 (linear) differential Pad\'e analysis on 1654 terms of 
the series (\ref{seriescompoH1H2})
gives a dominant singularity at $0.046028337184$ 
(in complete agreement with (\ref{seriescompoH1H2-xc}))
and, furthermore, gives an exponent equal to  $0$ corresponding 
to a logarithmic singularity. One also sees a number of 
singularities around $\, 1/16$, characteristic of an irregular 
singularity, see \ref{transcendantdiffPade}. 

Modulo $ 5$ the  series (\ref{seriescompoH1H2}) reads:
\begin{eqnarray}
\label{seriescompoH1H2mod5}
\hspace{-0.95in}&&   \quad  \,\,\,\,  
 H_1(H_2(x)) \,\, = \, \,\,\, 
1\,\, \,+4\,x\,\, +2\,{x}^{2}\,+2\,{x}^{3}\, +3\,{x}^{4}\, 
+2\,{x}^{5} \, +3\,{x}^{6}\, +3\,{x}^{8}\, +{x}^{9}\, +3\,{x}^{10}
 \nonumber \\ 
\hspace{-0.95in}&&   \quad  \quad \quad \,  \,  \, 
 +3\,{x}^{11}\,  +{x}^{12} \, +2\,{x}^{13}\, +{x}^{14}\, +3\,{x}^{15}
+3\,{x}^{16}\, +2\,{x}^{17} \,  +4\,{x}^{19}\, +2\,{x}^{20}\, +{x}^{22}
 \nonumber \\ 
\hspace{-0.95in}&&   \quad  \quad \quad \,   \,  \, 
+{x}^{23}  +4\,{x}^{24}+{x}^{25} \, +3\,{x}^{26} 
\,\,\,  \,+ \, \, \cdots 
\end{eqnarray}

The previous holonomic function $H_2(x)$ (which is 
also a diagonal of a rational 
function~\cite{Short})  reduces, modulo 
$ 5$, to a very simple algebraic function $A_2(x)$: 
\begin{eqnarray}
\label{H2mod5}
\hspace{-0.95in}&&   \,\,\,\,
H_2(x)  \mod \, 5 \,\,\, = \, \,\, A_2(x) 
\,\, = \, \,\, {{x} \over { (1+\, 4 \, x \, +x^2)^{1/4}}} 
 \, \,= \, \,  \,  \,\,
x\,\, \, +4\,{x}^{2}\,\,+{x}^{3} \,\,
+4\,{x}^{6}\,\,+{x}^{7}
\nonumber \\ 
\hspace{-0.95in}&&   \quad  \, \quad 
\, \,\,  +4\,{x}^{8} \,   +{x}^{11}\,+4\,{x}^{12}\,
\,+{x}^{13}\, +4\,{x}^{26}\,+{x}^{27} \,+4\,{x}^{28}\,
+{x}^{31} \, +4\,{x}^{32}  \,\,\,\, + \,\, \cdots 
\end{eqnarray}

Naively one might expect, modulo $ 5,$ the series 
(\ref{seriescompoH1H2}) for $ H_1(H_2(x))$, namely 
(\ref{seriescompoH1H2mod5}) to also be the same as the series 
expansion of  $ H_1(A_2(x))$ modulo $ 5$. Indeed, if one 
performs the series expansion of  $ H_1(A_2(x))$ modulo $ 5$, 
one gets
\begin{eqnarray}
\label{H2mod5serH1A2}
\hspace{-0.95in}&&   \quad 
H_1(A_2(x)) \,\, \,  = \, \,\, \,  
_2F_1\Bigl(\Bigl[{{1} \over {2}}, \, {{1} \over {2}}\Bigr],\, [1], \, 
16   \,  {{x} \over {(1+\, 4 \, x \, +x^2)^{1/4}}} \Bigr) 
 \\ 
\label{H2mod5ser}
\hspace{-0.95in}&&   \quad \, \,  \,\, = \, \,\, \, \, \,
1\, \,  \, +4\,x\,\, +2\,{x}^{2}\,+2\,{x}^{3}+3\,{x}^{4}
+2\,{x}^{5}\, +3\,{x}^{6} \,+3\,{x}^{8}\, +{x}^{9}\, +3\,{x}^{10} \,+3\,{x}^{11}
 \nonumber \\ 
\hspace{-0.95in}&&   \quad  \quad \quad  \quad \, \,
 +{x}^{12} \,\,+2\,{x}^{13}\,+{x}^{14}\,+3\,{x}^{15}\,
+3\,{x}^{16}\,+2\,{x}^{17} \,+4\,{x}^{19}\,+2\,{x}^{20}\,+{x}^{22}
\,+{x}^{23} 
 \nonumber \\ 
\label{H2mod5bis}
\hspace{-0.95in}&&   \quad  \quad  \quad  \quad \, \, 
+4\,{x}^{24}\,\, +{x}^{25} \, +3\,{x}^{26} 
\, \,\,  + \, \, \cdots,  
\end{eqnarray}
which is (as expected) the same expansion as (\ref{seriescompoH1H2mod5}).
However, recalling the previously mentioned closure properties of 
diagonals of rational functions by 
{\em right composition with algebraic functions}, 
the form $ H_1(A_2(x))$ in (\ref{H2mod5serH1A2}) can also be 
viewed as a diagonal of a rational function and, thus, the 
series (\ref{H2mod5ser}) should be the expansion of an 
algebraic function. Not surprisingly it is also the series 
expansion of $ A_1(A_2(x))$ modulo $ 5$.

We found the algebraic function satisfying, modulo $ 5$, 
the polynomial equation:
\begin{eqnarray}
\label{H2mod5}
\hspace{-0.95in}&&   \quad \quad  \quad  \quad 
({x}^{8} +{x}^{6} +4\,{x}^{5} +2\,{x}^{4}+3\,{x}^{3}+3\,{x}^{2}+3\,x+1)
 \cdot \, F(x)^{16} 
\nonumber \\ 
\hspace{-0.95in}&&   \quad  \quad \quad  \quad  \quad 
\, \, \, \, + \, ({x}^{4}+3\,{x}^{3}+3\,{x}^{2}+3\,x+1) 
 \cdot \,  (F(x)^{12}\, +F(x)^{4}\, +1)
  \\ 
\hspace{-0.95in}&&   \quad  \quad \quad  \quad  \quad \,\, \, \, 
+ \, ({x}^{2}-x+1)  \cdot \, (3\,{x}^{4}+5\,{x}^{3}+6\,{x}^{2}+4\,x+1) 
\cdot \,  F(x)^{8} 
 \,\, \,  \, = \, \,\,  \,  0.
  \nonumber
\end{eqnarray}

Of course there is nothing special about the prime $5$:
the series clearly reduces to 
algebraic functions modulo primes, or powers of a prime 
(see \ref{subclosureapp}).

The {\em other solutions} of the differentially algebraic 
equation (\ref{AH2mod5}) also reduce to algebraic functions 
modulo primes, or powers of a prime (see \ref{reducmodapp}).

Such simple results provide a strong incentive to  
systematically study the reduction, modulo primes, or powers 
of a prime,  of all the solution-series {\em with integer
coefficients}
of differentially algebraic equations, and more generally all 
{\em globally bounded} (\ref{G5}~\cite{Short}) solution-series 
of differentially algebraic equations. 

\subsection{Composition of holonomic functions are differentially algebraic. \\}
\label{compodiffalg}

Recalling the previous composition of two simple ${}_2F_1$ hypergeometric 
functions that are diagonals of rational functions, namely 
(\ref{compoH1H2}), it is  straightforward (but quite tedious), 
to find that the composition (\ref{compoH1H2}) of these two holonomic 
functions is {\em actually differentially algebraic}, 
and, more generally, that any composition
of holonomic functions is  differentially algebraic. This is 
a particular case of the more general result that the {\em composition 
of two differentially algebraic functions is differentially algebraic} 
(see~\cite{Bergeron}).  
In this example, denoting the series  (\ref{compoH1H2}) by 
$F(x) = H_1(H_2(x)),$ one finds that it satisfies 
a polynomial equation 
\begin{eqnarray}
\label{H2mod5pol}
\hspace{-0.95in}&&   \quad \quad \quad \quad \quad \quad  \quad  \quad 
P(x, \, F(x), \,F'(x), \,F''(x), \,F'''(x), \,F^{(4)}(x))
 \, \, \, =\,  \, \, \, 0, 
\end{eqnarray}
where the polynomial $P$ is the sum of  $9972 $ monomials 
of degree at most, $32$ in $x$,  $6$ in $F(x)$, 
  $11$ in $F'(x)$,  $11$ in $F''(x)$, 
$7$ in $F'''(x)$ and  $4$ in $F^{(4)}(x)$. This 
polynomial equation can be written as
\begin{eqnarray}
\label{H2mod5bis}
\hspace{-0.95in}&&   \quad \quad \quad \quad \quad \quad 
\sum_{n=0}^{32}  \, \,
   x^n \cdot \, P_n(F(x), \,F'(x), \,F''(x), \,F'''(x), \,F^{(4)}(x))
 \, \, \, =\,  \, \,\, \, 0, 
\end{eqnarray}
where the polynomials $P_n$ are homogeneous polynomials 
of degree $11$ in $F(x), F'(x), \cdots, F^{(4)}(x)$. If one
 scales $F(x)$ as $F(x) \rightarrow A \cdot  F(x)$,
(and consequently $F^{(i)}(x) \rightarrow A \cdot  F^{(i)}(x)$), 
the polynomial $P$ in (\ref{H2mod5pol})
scales as $P \rightarrow A^{11} \cdot P$. Therefore, if 
$ F(x)$ is a solution of (\ref{H2mod5pol}), 
$ A \cdot  F(x)$ is also a solution of (\ref{H2mod5pol}). 

This is a general result: for 
compositions of holonomic functions, the differentially
algebraic equation inherits the (differential Galois) symmetries 
of the linear differential operator corresponding to the first
 holonomic function in the composition (see \ref{compodiffalgapp}).

\vskip .1cm

\vskip .1cm

{\bf Remark 1:} The non-linear differential equation 
(\ref{H2mod5pol}) has, of course, many more solutions 
(see  \ref{Othersolapp1}) than the solution given by the 
series (\ref{seriescompoH1H2}) for which  one has a very precise 
location for one critical point namely the transcendental 
value $ x_c \simeq \, 0.04602833718455$,
see (\ref{seriescompoH1H2-xc}). It is clear that,
in contrast with  linear differential equations, where the 
(algebraic) 
{\em critical points correspond to the zeros of the head polynomial} 
of the linear differential operator, such a 
(transcendental) critical value (\ref{seriescompoH1H2-xc}) 
{\em cannot be simply seen in the non-linear ODE} 
(\ref{H2mod5bis}). The singularities of 
the solutions of (\ref{H2mod5bis}) depend on the ``initial 
conditions'', the first terms of the solution series:
these singularities are {\em movable} singularities 
(see \ref{Othersolapp1}).

The occurence of {\em movable singularities} 
(that can be infinite in number ...) is a phenomenon 
that we will see occur systematically in almost 
all the differentially algebraic equations emerging 
in our studies.

Along this line it is interesting to remark that the traditional 
diff-Pad\'e analysis performed on so many series of lattice 
statistical mechanics or enumerative combinatorics, tries 
to associate to a given series the best {\em linear} 
differential equation with {\em polynomial coefficients}, which 
{\em necessarily has fixed singularities}. With this 
differentially algebraic example (\ref{compoH1H2}), which requires
a non-linear differential equation (\ref{H2mod5bis})  with movable 
singularities to be correctly described, it is interesting to see
the kind of results and bias that are obtained from the 
traditional diff-Pad\'e analysis forcing, by construction, 
fixed singularities, on this movable singularities example. This 
is sketched in \ref{transcendantdiffPade}. 

 With the ``toy example'' (\ref{compoH1H2}), 
where  one has a very precise 
location for one critical point,
 it is interesting to see what kind of results a 
traditional {\em linear} differential Pad\'e analysis would 
give on such a {\em non-linear} differentially algebraic series 
(\ref{compoH1H2}). In a composition of two holonomic functions 
$\, H_1(H_2(x))$, one might expect, in general, that the 
critical points that are not algebraic numbers anymore~\cite{Short}, 
but {\em transcendental numbers}, such that $\, H_2(x)$
is a singular point of $\, H_1(x)$, are, nevertheless all regular 
singular points, their critical exponents  being simply related 
to the ones of the $ _2F_1$ hypergeometric series. The singular 
points of $\, H_2(x)$ may yield irregular singularities 
for $\, H_1(H_2(x))$.

\vskip .1cm

\subsection{Compositions of  functions. \\}
\label{composifunct}

\vskip .1cm

 From these simple examples we see that 
the {\em composition} of diagonals of rational functions, 
 or even {\em algebraic functions} of diagonals
of rational functions\footnote{In fact there is no reason 
to restrict ourselves to the composition of holonomic 
functions, like diagonals of rational functions. One can also 
compose algebraic functions of diagonals of rational functions 
(see (\ref{moreinvolvedalg})).
}, build an extremely large class of functions:
their series are globally bounded and, so, can be recast 
to have {\em integer coefficients}, 
 a {\em non-zero radius of convergence}, and will
{\em reduce to algebraic functions modulo every prime and power of a prime}, 
these functions {\em being necessarily differentially algebraic}.

One probably has a much more general result: 
if two series with integer coefficients 
are such that they reduce to algebraic functions modulo every $p^r$, 
so does the composition of these two series\footnote{If the two series
also have a non-zero radius of convergence so does their composition.}.
Note however that this class  {\em has no reason 
to be differentially algebraic}.
Actually, if two series $S_1$ and  $S_2$ with integer coefficients
({\em not necessarily differentially algebraic}) reduce 
to algebraic functions  $A_1$ and  $A_2$ 
modulo every prime, or power of a prime,
the composition $S_1(S_2(x))$ reduces to  $A_1(A_2(x))$ 
modulo that prime, or power of a prime.

It is natural to ask if the class of arbitrary compositions 
of algebraic functions of diagonals of rational functions is 
sufficiently large to exhaust all  series 
with integer coefficients, with non-zero radius of convergence,  
reducing modulo every prime (or power of a prime) to an algebraic 
function? This is, and this will probably remain  for some 
time, an open question. 

\vskip .1cm

For such a large class of functions one can imagine that 
the reduction to algebraic functions modulo a prime can be 
difficult to see, the polynomial encoding this  algebraic 
function being of {\em very high degree} (see \ref{AFD} 
and \ref{subclosureapp} in a differentially algebraic framework). 
 Is this perhaps the situation we encounter,
modulo $ 3$, $ 5$, etc.  for the full susceptibility 
of the Ising model? 

\vskip .1cm 

Since this very large class of functions obtained by compositions 
of an arbitrary number of diagonals of rational functions, 
or more generally, algebraic functions of diagonals of rational 
functions, actually corresponds to differentially algebraic 
functions, we have some incentive to see if the full 
susceptibility of the square-lattice Ising model is a 
differentially algebraic function. 
 
\vskip .1cm

\section{Getting $ 5043$ terms for the susceptibility series generation.}
\label{IwanGetting}

In earlier work~\cite{CGNP11}, we described the generation and 
analysis of the Ising susceptibility series to $2048$ terms\footnote{
Similar large series  were also obtained  and analysed
with a totally different corner transfer matrix approach for the 
triangular and honeycomb lattices but only to 600 terms~\cite{CGNP11}. 
The singular part of the susceptibility was, however, obtained to 
great precision~\cite{CGNP11}. } on the square lattice 
in natural high- and low-temperature variables. By making some 
minor improvements to the program, and running for longer 
on a larger machine, we have extended the square-lattice 
series to 5043 terms~\cite{Iwan}. 

These very long series confirm all the reductions 
to algebraic functions modulo $2^r$ obtained in~\cite{Auto}. 
These new data would also allow the precision of the earlier analysis
to be greatly enhanced.
However we consider that there is little benefit in doing so, 
as the existing precision is so far in advance of any theoretical 
or experimental application that there is nothing to be gained 
in pursuing this (local) scaling approach.

As discussed in~\cite{ONGP01} and~\cite{Nickel99}, the square-lattice 
Ising susceptibility is believed 
to have a {\em natural boundary} on the unit circle in the complex 
$s/2= \sinh{2K}/2$ plane and thus cannot be holonomic. Even with very 
long series it is extremely difficult to see this fundamental 
{\em natural boundary} property. Seizing the opportunity of the 
{\em polynomial-time complexity of the algorithm}~\cite{ONGP01}, 
the motivation to get even longer series for the Ising susceptibility 
cannot be justified by a desire to extend such a (local) 
scaling analysis, or even an attempt 
to better understand the complex analytical structure of 
the susceptibility. In contrast the results obtained 
in~\cite{Auto} corresponding to the reduction modulo 
$2^r$ of the susceptibility series to algebraic 
functions, are of a more {\em arithmetic and global character}, 
and fully justify efforts to get series expansions with 
more than 2048 terms. Rather, we seek to investigate 
the (global) nature of the susceptibility, as revealed 
by the behaviour of the coefficients modulo a prime, or  
power of a prime~\cite{Auto}. 

The program, which makes use of Fast Fourier Transform methods 
for multiplications, can be sped up essentially linearly with 
the number of processors. The series were calculated modulo 
many different primes  of the form $p_k=2^{15}  -r_k$ and
the exact integer coefficients were then obtained from the set of 
remainders using the Chinese remainder procedure.
The series coefficients $c_n$ grow
 asymptotically as $4^n$  so to get a series with $N$ terms
requires approximately $2N/15$ primes. 
The calculation took some $ 22.5$ hours per prime\footnote{On 
a newer machine the calculation per prime takes $ 12$ hours,
the $ 5043$ terms calculation taking close to 6000 hours in total.}.
In order to obtain 5043 terms~\cite{Iwan} of the 
(low and high-temperature) series  we used 700 primes
 and the total
 calculation  therefore required close to 16000 hours in total.  
This has been achieved using 700 processors
of the National Computational Infrastructure (NCI)  
facility at the Australian National University. 
The parallel version of this program
just runs several primes simultaneously. 

\vskip .1cm

{\bf Remark 1:} It is tempting to imagine further extending 
these series, restricted modulo particular integers, for 
instance powers of $2$, to  further check the reduction of 
the susceptibility series to algebraic functions modulo 
$ 2^r$. The current version the program performs 
{\em integer divisions}. This is only possible when one 
performs the calculation {\em modulo a prime number}. 
Therefore one {\em cannot calculate 
modulo powers of $ 2$ directly}: one needs to calculate 
the series exactly (in characteristic zero) 
as has been done to 5043 terms and then from the exact 
series coefficients one can of course do any modulo integer 
calculations one wishes. Unless one can re-write the 
algorithm to avoid integer division there is little prospect 
of pushing these series modulo powers of a prime much further.

\vskip .1cm

{\bf Remark 2:} The nature of the algorithm unfortunately is such 
that one {\em cannot run it modulo a small prime} 
(like $ 3$, $ 5$, ...). The prime 
{\em must be larger than the length of the series}, 
due to the integer divisions that take place. In principle, one could 
calculate 10000 terms modulo a single prime greater than 10000. 
Unfortunately, when non-holonomic series reduce to algebraic functions 
modulo a prime $ p$, we have seen~\cite{Selected} that the complexity 
of the corresponding polynomial (encoding the algebraic function)
grows with the prime $ p$. The polynomials corresponding to these
very long 10000 terms series would be huge, the 10000 terms 
being insufficient to find the polynomial (if they exist). 

\vskip .1cm

{\bf Remark 3:}  To get longer series would require one to perform 
the Chinese reminder procedure with more primes (each prime
being larger than the length of the series). 
If we want to get 10000  terms for the susceptibility 
series, we would need to perform the calculations 
modulo more than 1350 different primes of 
the form $ 2^{15}-r_k$. Naturally if one used 
primes of the form say  $2^{30}-r_k$
one would need half as many primes. However, in that case
one would need to do integer multiplications using 64-bit integers.
Furthermore the FFT multiplication of series is done using
double-precision floating-point numbers (the remainders 
of the integer coefficients modulo the prime recovered 
from the floating-point numbers). The use of larger primes 
may require the use of higher precision floating-point calculations 
(such as quad-precision) in order to correctly reproduce the
integer valued remainders.  We note that a 
calculation of the 10000 term series took almost 68 hours 
(on a Mac Pro desktop 
with a 3.5 GHz 6-Core Intel Xeon E5 processor).
So to get the exact series to order 10000 would require 
at least 90000 CPU hours
on this machine. 
  
\vskip .1cm

\subsection{The  $5043$ terms series modulo $3, 5, 7, \ldots \, $: 
no reduction to algebraic functions? }
\label{noreduction}

Now that we have such long series for the susceptibility, we can 
revisit the results given in~\cite{Auto}, namely that the (low or
high temperature) susceptibility 
series reduce to algebraic functions modulo $ 2^r$.  This is 
verified immediately. For instance, we find with $5043$ terms
of the high-temperature full susceptibility series $ \chiH(w)$, 
the simple functional equation modulo $ 32$ 
(see equation (47) in~\cite{Auto}):
\begin{eqnarray}
\label{tildechi3func16}
\hspace{-0.95in}&&  \,\,\,      \quad     \quad     \quad     \quad  
 \chiH(w^2) \,\,\, = \,\,\,\,\,
 w \cdot \,  \chiH(w)
 \, \,\, \,  + \, \,  8 \, w^3 \cdot \, (2\,w^{15} \, -w^7 \,+w \,-5 ). 
\end{eqnarray}
Modulo $64$,  $128$, one verifies~\cite{Auto} with $5043$ 
terms that one cannot distinguish between $  \chiH$ and 
$ \tilde{\chi}_H^{(1)} +  \tilde{\chi}_H^{(3)} +  \tilde{\chi}_H^{(5)}$, 
which is the diagonal of a rational function~\cite{Short}, and 
{\em thus reduces to an algebraic function} modulo any prime or power 
of a prime, and in particular modulo $ 64$,  $ 128$. Similarly, with 
the $ 5043$ terms of the low-temperature series of the full 
susceptibility $ \chiL$, one verifies~\cite{Auto}
that one cannot, modulo 
$ 2,  4,  8,  16,  32,  64,  128,  256$,
 distinguish between $ \chiL$ and 
$ \tilde{\chi}_L^{(2)}  +  \tilde{\chi}_L^{(4)} +  \tilde{\chi}_L^{(6)}$
which is  the diagonal of a rational function~\cite{Short}, and 
{\em thus reduces, modulo} $ 2,  4,  8,  16,$
$  32,  64,  128,  256$, 
{\em to an algebraic function}. 
{\em All the exact results of}~\cite{Auto} 
{\em remain valid with $ 5043$ terms}. 

In contrast, the $ 5043$ terms 
susceptibility series {\em does not seem} to reduce 
to algebraic functions even modulo $ 3$, $ 5$, $ 7$, $ 11$, 
etc. This is quite a puzzling result, when one compares
this negative result with the comparatively easy way~\cite{Auto} 
reduction to algebraic functions are obtained modulo a large set 
of powers of $ 2$, and also the reasonably easy way~\cite{Selected}
reductions to algebraic functions are obtained for the
solutions of Tutte's equation (\ref{Tutteq}), and the 
miscellaneous examples discussed in \ref{appendix3}.

At this stage it is not possible to conclude that this negative 
result with $ 5043$ terms really means that the susceptibility series 
does not reduce to algebraic functions modulo $p^r$ for $p \neq 2$. 
It could be that the algebraic 
functions modulo $ 3$, $ 5$, $ 7$, $ 11$, 
etc are drastically more complicated, so that even a $ 5043$
 term series is too short.

With the simple quadratic example (\ref{roots}) 
of Section (\ref{moreinvolvedalg}) we showed that this 
{\em algebraic function of the diagonal of a rational function} 
 reduces modulo $p^r$ to an algebraic function. However, even 
in this very simple  example, the corresponding polynomial 
is already, modulo $ 7$, of degree $ 60$ in $ x$ and $ 72$ 
in $ z$ (see \ref{AFD}). Switching 
from ratios of diagonals of rational functions (see Paragraph 5.2 
of~\cite{Selected})  to  algebraic functions  of diagonals of 
rational functions may result in drastically
more involved  algebraic functions for the reductions modulo primes.

We found similar results for the composition of very simple holonomic 
functions of Section (\ref{composidiag}), 
and their reduction modulo $ 7$
(which gave a polynomial equation of degree $ 36$ in $ z$ 
and $ 18$ in $x$, see (\ref{v1mod7}) in \ref{subclosureapp}). 
These two examples (algebraic 
expressions of diagonals of rational functions
and compositions of diagonals of rational functions) were extremely
simple pedagogical examples of differentially algebraic functions
(see (\ref{roots})  and (\ref{compoH1H2})). It is 
quite natural to imagine that, if the square-lattice Ising 
susceptibility is indeed a differentially algebraic function, it 
could correspond to quite a large
polynomial differential equation (much more involved than Tutte's 
example (\ref{Tutteq}) of Section \ref{Tutte}), and that finding 
the polynomial relations associated with the reduction of the 
solution-series modulo primes and powers of primes would thus 
be quite difficult, their degree being very large. 

\vskip .1cm 

Along  similar  lines, it is worth recalling the 
hypergeometric example 
${}_3F_2([1/9, 4/9, 5/9],[1/3, 1], 3^6 x)$ 
introduced by G. Christol~\cite{Short}, 
a few decades ago, as an example of a holonomic 
$G$-series with {\em integer coefficients}\footnote[1]{The
fact that the corresponding series expansion is a series 
with integer coefficients is far from obvious 
(see appendix D in~\cite{Short}).} that 
{\em may not be the diagonal of a rational function}: after 
all these years, it is still an open question whether this 
function is, or is not, the diagonal of rational function!
In such cases it is not guaranteed that the corresponding series 
modulo primes, or powers of primes, are algebraic functions.
The analysis, modulo primes, and powers of primes, of such 
an example of a holonomic function is quite puzzling:
in~\cite{Selected} it was 
shown (see Appendix B.4 in~\cite{Selected}),
that it becomes {\em extremely difficult to see whether 
a series like ${}_3F_2([1/9,4/9,5/9],[1/3,1],3^6 x)$, 
is, modulo primes, an algebraic function} or not, even for 
small prime numbers!
This may be seen as a consequence of the fact that 
 such a holonomic function is 
``extremely reluctant'' to be seen as the diagonal of a rational 
function. Since we want to see {\em differentially 
algebraic functions as a natural generalisation of diagonals of 
rational functions}, we need to keep in mind that simple hypergeometric 
example: the difficulty we encounter in seeing the
full susceptibility series reducing to algebraic functions
modulo $3$ (or modulo $5$, etc) could be a
similar ``reluctance''. 

\subsection{The  $5043$ terms series: 
no differentially algebraic equation? }
\label{nodiffalg}

Let $F(x)$ be a generating function for which we have some finite number 
of known initial terms. Our algorithm searches for a polynomial
\[
	P(x, \, y_0,\, y_1, \,y_2, \,\ldots,\, y_k)
\]
such that
\begin{equation}
	\label{suchthat}
	P(x, F(x), F'(x), F''(x), \ldots, F^{(k)}(x)) = 0.
\end{equation}
Let $Q$ denote the left-hand side of the above equation. Our 
algorithm constructs a variety of potential forms $Q$ and 
for each checks if the known initial terms satisfy a 
functional equation of that form. Each possible $Q$ 
is a sum of terms of the form
\[
	p_i(x) \cdot \, (F(x))^{c_i(0)} \cdot \, 
   (F'(x))^{c_i(1)} \, \cdots \, (F^{(k)}(x))^{c_i(k)},
\]
where $p_i(x)$ is called the \emph{polynomial coefficient} and 
the rest is called the \emph{functional term}. Clearly, 
each possible functional term can be thought of as a 
weak composition (that is, its summands may be zero), 
$c_i = (c_i(0), c_i(1), \ldots, c_i(k))$, and so each form $Q$ 
is associated with some set of compositions, along with the 
degrees of the polynomial coefficients.

Each form $Q$ to which we attempt to fit our known initial terms is uniquely 
specified by a $3$-tuple $(m,k,d)$. Here, $d$ is the maximum degree of 
all polynomial coefficients. The set of functional terms corresponds 
to the set of all compositions of $m$ into exactly $k+1$ parts. 
For example, $(m,k,d) = (3, 1, d)$ corresponds 
to the set of compositions
\[
	\{(0,0), \, \, (1, 0),\, \,  (0, 1), \,  \,(2, 0), \,  \,(1, 1), \, \, (0, 2), \,  \,(3, 0), \, \, (2, 1),\,  \, (1, 2), \, \, (0, 3)\},
\]
and thus the functional terms of this particular $Q$ are
\[
	\begin{array}{lllll}
		1 \qquad\qquad & F(x) \qquad & F'(x) \qquad \qquad & F(x)^2 \qquad \qquad & F(x)F'(x)\\
		F'(x)^2 & F(x)^3 & F(x)^2F'(x) & F(x)F'(x)^2 & F'(x)^3.		
	\end{array}
\]

The quantity $m$ is called the \emph{algebraic order}, while $k$ is the 
\emph{differential order}. Using the known initial terms for $F(x)$, 
we calculate the functional terms of each component of the predicted 
form. The polynomial coefficients represent the unknowns to be 
determined. Each has the form
\begin{eqnarray}
\label{pi}
\hspace{-0.95in}&&  \,\,\,      \quad     \quad   \quad   \quad     \quad
	p_i(x) \,\,   = \, \, \, 
   a_{i,d} \cdot \, x^d \, + a_{i,d-1}\cdot \, x^{d-1} \, 
+  \, \cdots \, + a_{i,1}\cdot \,x \, + a_{i,0},
\end{eqnarray}
and therefore if the predicted form has $R$ components, the total number of 
unknown quantities is $R\cdot(d+1)$. As we are seeking a polynomial $P$ 
such that $P(x,F(x), \cdots, F^{(k)}(x)) = 0$, the predicted form can be 
expanded to give a series in $x$ in which each coefficient must equal 
zero. So long as a sufficient number of initial terms of $F(x)$ are 
provided, we can attempt to solve the system. When the system involves 
$R \cdot(d+1)$ unknowns, we provide $R\cdot(d+1) + T$ equations in these 
variables--we typically use $T=10$. This provides some confidence that 
a conjectured algebraic differential equation is in fact satisfied 
by the unknown series. 

The procedure to check whether the known initial terms of a series satisfy 
an algebraic differential equation now proceeds as follows. Given $N$ 
known initial terms, we first calculate 
all tuples $(m,k,d)$ for which the number of unknowns $U$ is at 
most $N - k - T$. Each time a derivative of the known initial terms 
is computed, the number of unknowns needed increases by one. 
This is why we require $U + k \leq N-T$.
Some of these predicted forms are properly contained in others, 
e.g. there is no need to check for a predicted form 
$(m,k,d) = (3, 1, 5)$ if we intend to also check $(m,k,d) = (3, 2, 7)$. 
Therefore, we filter out all such redundancies. What remains 
is a list of maximal predicted forms for which there are a 
sufficient number of known initial terms to check. For each, 
we construct the components, extract the 
linear system, and attempt to find a solution.

To demonstrate the method, consider the sequence defined by
\begin{eqnarray}
\label{hnplus2}
\hspace{-0.95in}&&  \,\,\,      \quad \quad \quad \quad \quad \quad     \quad
	h_{n+2} \,  \,  =  \, \,  \, 
 \sum_{i=1}^{n} \,   i \cdot  \, h_{i+1} \, h_{n-i+2},
\end{eqnarray}
with initial conditions $h_0 = h_1 = 0$ and $h_2 = 1$ (see \ref{appendix32} 
for more about this sequence). 
The first few term of the power series of this sequence are
\begin{eqnarray}
\label{hnplus2}
\hspace{-0.95in}&&  \,\,\,      \quad \quad  
\quad \quad \quad     \quad
	F(x)  \, = \,  \,   \, 
x^2  \, + x^3  \, + 3x^4  \, + 14x^5  \, + 85x^6  \,  \, + \cdots,
\end{eqnarray}
and it is computationally easy to get thousands of known initial 
terms. If we provide at least $35$ initial terms to our algorithm 
(using $T=10$), it eventually reaches the form $Q$ corresponding to 
$(m,k,d) = (2, 1, 3)$. This corresponds to checking if $F(x)$ 
satisfies a functional equation of the form
\begin{eqnarray}
\label{infinitesum}
\hspace{-0.95in}&&  \, \,    \quad    \quad  \quad \quad   \,  \,
	p_0(x)  \, \, + p_1(x)\cdot  \, F(x) \, \,  
 + p_2(x)\cdot  \,F'(x) \,\,   + p_3(x)\cdot  \,F(x)^2 
\nonumber \\
\hspace{-0.95in}&&  \, \,   \quad \quad  
\quad  \quad \quad  \quad    \quad   \, \,
    \, + p_4(x)\cdot  \,(F'(x))^2 \,  \, 
+ p_5(x)\cdot  \,F(x)\, F'(x) \, \,  = \, \,\,  0,
\end{eqnarray}
where each $p_i(x)$ has degree at most $3$. Into this equation we 
substitute the first $35$ terms of the power series of $F(x)$. Upon 
evaluating $F'(x)$ and computing the product $F(x)F'(x)$, 
we are left with a power series on the left-hand side involving the 
unknown coefficients of each of the polynomials $p_i(x)$. As each term 
of this power series must equal zero, we can attempt 
to find a solution for the unknown coefficients of the $p_i(x)$ 
by solving a linear system where each equation is a coefficient 
of the left-hand side. This system has $34$ 
equations and $24$ unknowns, and has a solution corresponding to
\[
	p_0(x) = x^3, \quad \, p_1(x) = -x, \quad \, p_2(x) = 0,
       \quad \, p_3(x) = -1, \quad \, p_4(x)=0, \quad \, p_5(x) = x.
\]
Therefore, we have used $35$ initial terms of $F(x)$ to conjecture 
that $F(x)$ satisfies the equation
\[
	x^3 \, - x \cdot \, F(x) \,\, - F(x)^2\, 
                  + x \cdot \,F(x)\, F'(x) \, \, = \,\, \, 0.
\]
Our implementation performs all of these calculations in under a second.

\vskip .1cm

\subsection{Running the program: some simple examples.}
\label{running}

\vskip .1cm

\subsubsection{Running the program: a first simple pullback hypergeometric example. \\}
\label{running1}

We first consider a simpler example of composition than that
of two ${}_2F_1$ hypergeometric functions given in 
Section (\ref{composidiag}): rather we consider the composition 
of a ${}_2F_1$ hypergeometric function with $x\ln(1+x)$:
\begin{eqnarray}
\label{considersimpler}
\hspace{-0.95in}&&   \quad   \quad \qquad \qquad 
   F(x) \,  \, = \, \,  \, \, 
  _2F_1\Bigl( \Bigl[{{1} \over {2}}, \, {{1} \over {2}}\Bigr], \, [1], 
\, 16 \, x    \ln(1+x) \Bigr).
\end{eqnarray}
This function has an irregular singularity at $\, x\, = \, -1$.
Using our program with more than $2407$
coefficients  one can find (in 4834 seconds) 
one non-linear differential equation (\ref{suchthat}) 
for this differentially algebraic function, 
of {\em order three}, of degree $18$ in $x$, and
degree $5$ in $F(x)$, $F'(x)$, $F''(x)$ and  $F'''(x)$.
Note that  
\begin{eqnarray}
\label{considersimplerother}
\hspace{-0.95in}&&   \quad   \,  \,  \,  \, 
   F(x) \,  \, = \, \,  \,  \, 
 _2F_1\Bigl( \Bigl[{{1} \over {2}}, \, {{1} \over {2}}\Bigr], \, [1], 
\, \,  16 \, x    \ln(1+x) \, + \, 16 \,\alpha   \, x\Bigr)
\nonumber \\ 
\hspace{-0.95in}&&   \quad   \quad   \quad 
\, = \,\, \,\,  1 \,\, \, +4\,\alpha  \cdot    \, x\, \,
+ \, (36\,{\alpha}^{2}+4)  \cdot    \, {x}^{2}\, \,
+ \, (400 \,{\alpha}^{3}+72\,\alpha-2)  \cdot  \, {x}^{3}\,\,\, + \, \, \cdots
\end{eqnarray}
is also a solution of the same non-linear differential equation 
(\ref{suchthat}). This one-parameter family of solutions 
(\ref{considersimplerother}) is a consequence of the fact that, for 
compositions of holonomic functions,  there always exists a non-linear 
differential equation (\ref{suchthat}) that 
inherits
the (differential Galois group)
symmetries of the underlying holonomic functions -- see
 for instance~\cite{Cox}.
This one-parameter family of solutions (\ref{considersimplerother})
clearly illustrates that the solutions of the 
associated non-linear differential equation (\ref{suchthat})
have {\em movable singularities} corresponding to 
$ \,16 \, x  \,  \ln(1+x) \, +  16 \,\alpha  \,  x  = \, 1$
which  depend on the parameter $ \alpha$ of the 
{\em initial conditions of the series} 
(\ref{considersimplerother}). 

\subsubsection{Running the program: a second simpler example of composition of holonomic functions. \\}
\label{running2}

We further illustrate these questions of symmetries, differential 
Galois groups, families of solutions and {\em movable singularities}
with another simple example of the composition
of holonomic functions $F(x)=  f(g(x)),$ where
\begin{eqnarray}
\label{extremely}
\hspace{-0.95in}&&   \, 
f(x) \, = \, {{1} \over {1\, -x}}, \quad \, \, 
g(x) \, = \, \, x    \ln(1+x), \quad \, \, \,  \,
\, \, f(g(x)) \,  = \, \,  
{{1} \over { 1 \, - \, x   \ln(1+x)}}.
\end{eqnarray}
The two functions $f(x)$ and $g(x)$ are solutions of 
 order-one and  order-two linear differential operators,
$(x-1)\,  D_x \,+1$ and 
$(x+2)\, -x\, (x+2)\, D_x\, +x^2\, (1+x)\, D_x^2$,
which have, respectively, the one and two-parameter 
families of solutions:
\begin{eqnarray}
\label{family}
\hspace{-0.95in}&&   \, \, \, \, \,  \,  \quad  \quad 
f(\alpha ; \, x) \, = \,\, \,  {{ \alpha } \over {1\, -x}}, 
\quad  \quad \quad \,
g(\beta,\gamma ; \, x) \, = \, \,\, \, 
 \beta  \cdot \, x \cdot \, \ln(1+x) \,\,  + \gamma  \cdot \,x. 
\end{eqnarray}
Our program very quickly gives a non-linear differential equation
${\cal N}_1 = 0$, 
for $F(x) = f(g(x))$ given by (\ref{extremely}), 
where ${\cal N}_1$ reads:
\begin{eqnarray}
\label{extremelyequ}
\hspace{-0.95in}&&   \, \, \, \quad  \, \, \, \,  \, \,   \, \,
 {\cal N}_1 \,  \, = \, \, \, \,
  (1+x) \cdot \,F(x) \,\, 
\,\, -(x^2+x+1) \cdot \, F(x)^2 \, \, 
+ \, x \cdot \, (1+x) \cdot \, F'(x),
\end{eqnarray}
and one readily verifies that 
\begin{eqnarray}
\label{f11gamam}
\hspace{-0.95in}&&   \, \, \,  \quad  \quad  \,  
f(1; \, g(1,\gamma ; \, x)) 
\, \,\, = \, \,\,\,  \,  {{ 1} \over {
1 \, - \, x \cdot \, \ln(1+x) \, - \gamma  \, x }}
\nonumber  \\
\hspace{-0.95in}&&   \, \, \,  \quad   \quad   \quad  \quad  
\,   \, \,\, = \, \,\,  \, \,\,  
1 \, \, \,  +\gamma \cdot \,x \, \,  \, 
+ \, ({\gamma}^{2}+1) \cdot \, {x}^{2} \, 
+ \left( 2\,\gamma+{\gamma}^{3} -{\frac {1}{2}}\right) \cdot \, {x}^{3} 
\,\,   \, \, + \, \, \cdots 
\end{eqnarray}
is a solution of (\ref{extremelyequ}). 

Note that 
\begin{eqnarray}
\label{extremelyabc}
\hspace{-0.95in}&&   \quad   \quad  \qquad 
   F(x) \,  \,  = \, \, \, \, f(\alpha ; \, g(\beta,\gamma ; \, x))
 \,\,\, = \, \, \, \,
 \,  {{\alpha} \over {
1 \, -\, \beta  \, x \cdot \, \ln(1+x) \, - \gamma  \, x}}.
\end{eqnarray}
is {\em not} a solution of (\ref{extremelyequ}) in general: 
it is a solution {\em only when}  $\beta = \alpha = 1$.

Actually, if one looks for the power series solutions of 
the non-linear differential equation (\ref{extremelyequ}),  
$F(x) = a_0 + a_1 x + a_2 x^2 + \cdots$,
one finds immediately the condition $a_0 (a_0 \, -1) =  0$.
If $a_0  =  1$ one gets, order by order, the one-parameter 
series (\ref{f11gamam}), when $a_0 = 0$ gives the null 
function $F(x) = 0$.

There is another non-linear differential equation 
${\cal N}_2 = 0$, where ${\cal N}_2 $ reads
\begin{eqnarray}
\label{otherequ}
\hspace{-0.95in}&&   \, \, \,    \quad  \,  
{\cal N}_2 \, = \, \, \,
(x+2) \cdot \, F(x)^2 \, \,  \,
+ x  \cdot \, (x+2)  \cdot \, F(x) \cdot \, F'(x)\,  \,\,
 \\
\hspace{-0.95in}&&   \, \, \, \quad  \,   \quad 
 +2  \cdot \, (1+x)  \cdot \, (x^2+x+1)  \cdot \, F'(x)^2  
\,\,\, -(1+x)\cdot \, (x^2+x+1) \cdot \, F(x) \cdot \, F''(x),
\nonumber 
\end{eqnarray}
which is {\em homogeneous} (quadratic) in 
$F(x)$, $F'(x)$ and $F''(x)$. This homogeneity 
is inherited from the (differential Galois group) symmetry 
of the order-one operator $(x-1) \,D_x \,+1$. 
One readily verifies that 
 $f(\alpha ; g(1,\gamma ; x))$
is  {\em also}  a solution of (\ref{otherequ}).
Again if one looks for the power series solutions of the 
non-linear differential equation (\ref{otherequ}), one 
finds, order by order, that either $F(x) = 0$ or 
$F(x)  = \, f(\alpha ; g(1,\gamma ; x))$.

There is a third non-linear differential equation 
${\cal N}_3 = 0$, where ${\cal N}_3$ reads:
\begin{eqnarray}
\label{anotherequ}
\hspace{-0.95in}&&   \, \, \,\,  \quad  \, 
{\cal N}_3 \, = \, \, \,
\, \,   (x+2)  \cdot \, (F(x)\,  - \, 1)  \cdot \, F(x)^2 
\, \, \,  \,  -x \cdot \, (x+2) \cdot  \, F(x) \cdot \, F'(x) \,
\nonumber \\
\hspace{-0.95in}&&   \, \, \, \quad \quad \quad  \quad
 \,\,-2\, x^2 \cdot \, (1+x)  \cdot \, F'(x)^2 \, \,\,
\,\, 
\, +x^2 \cdot \, (1+x) \cdot  \, F(x) \cdot \, F''(x) ,
\end{eqnarray}
which has as a solution  
$f(1 ; g(\beta,\gamma ; x))$. 
This time the (non-homogeneous) non-linear differential 
equation (\ref{anotherequ}) inherits  the 
(differential Galois group) symmetry of the order-two 
operator $(x+2)\, -x\,(x+2)\,D_x \, +x^2 \,(1+x)\, D_x^2$.
If one looks for the power series solutions of the non-linear 
differential equation (\ref{anotherequ}), one immediately 
obtains the condition $\, a_0^2 \, (a_0 -1) = \,0$. The 
 condition $a_0 = 0$ yields, order by order, $F(x) = 0$.
The  condition $a_0 = 1$ yields, order by order, the two-parameter
family of solutions  $F(x) =  f(1 ; g(\beta,\gamma ; x))$.

Finally there is a last  non-linear differential equation 
$ {\cal N}_4  =   0$, where $ {\cal N}_4$ reads:
\begin{eqnarray}
\label{lastequ}
\hspace{-0.95in}&&   \, \, \,  \quad    \, \,\, 
{\cal N}_4 \, = \, \, \,
6 \cdot \,{x}^{2} \cdot \, (x+2)  \, (1+x) \cdot \,  F(x)^{2} \, F'(x)^{3}
\, \, \,  -2\, {x}^{2} \cdot \, (x+3) \cdot \,  F(x)^{3} \, F'(x)^{2}
\nonumber \\ 
\hspace{-0.95in}&&   \, \, \,  \quad  \quad   \, 
\, -6 \cdot \,{x}^{2} \cdot \, (x+2) \cdot \, (1+x) 
\cdot \, F(x)^{3} \, F'(x)\, F''(x)\, 
\, \, +{x}^{2} \cdot \, (x+3) \cdot \, F(x)^{4} \, F''(x)
\nonumber \\ 
\hspace{-0.95in}&&   \, \, \,  \quad   \quad  \, 
\, +{x}^{2} \cdot \, (x+2)  \cdot \, (1+x) \cdot \, F(x)^{4} \, F'''(x),
\end{eqnarray}
which is {\em homogeneous} (quintic) in 
$F(x)$, $F'(x)$, $F''(x)$ and $F'''(x)$
and only this non-linear differential equation
has the three-parameter family (\ref{extremelyabc})
of solutions inherited from the one and two-parameter families 
of solutions (\ref{family}) of the order-one and order-two linear 
differential operators (i.e. inherited from the differential Galois 
groups of the order-one and order-two linear differential operators). 
If one looks for the power series solutions of the non-linear 
differential equation (\ref{lastequ}), one immediately obtains
the two conditions $a_0 = 0$ and 
 $2\, a_0^2 a_3\, +a_2 a_0\, (a_0 -4a_1)\,
  +a_1^2 \, (2 a_1  - a_0)  =  \, 0$. Condition $ a_0  =  0$ gives 
the null function $ F(x)  =   0$, when the second condition gives, 
order by order, the three-parameter series (\ref{extremelyabc}).

\vskip .1cm 

There are some relations between the non-linear ODEs, for instance:
\begin{eqnarray}
\label{relationsbetween}
\hspace{-0.95in}&&   \, \, \,  \quad  \quad  
{\cal N}_3 \, \,  = \,  \, \, \, 
x \cdot \, F(x) \cdot \, {{d {\cal N}_1} \over {dx}} 
\,\,  \,  -2 \cdot \, 
\Bigl(F(x) \, + \, x \cdot F'(x)  \Bigr) \cdot  {\cal N}_1,
\\
\hspace{-0.95in}&&   \, \, \,  \quad  \quad  
x \cdot {\cal N}_2 \, \,  = \,  \, \, \, 
-\, (x^2 +x +1) \cdot \, F(x) \cdot \, {{d {\cal N}_1} \over {dx}} 
\nonumber \\
\hspace{-0.95in}&&   \, \, \,  
\quad  \quad  \quad  \quad  \quad   \quad  \quad  
\,  \, 
+ \Bigl( (2\, x+1) \cdot \, F(x) \, 
+ \, (x^2 +x +1) \cdot \, F'(x) \Bigr) \cdot {\cal N}_1.
\end{eqnarray}
or
\begin{eqnarray}
\label{relationsbetween}
\hspace{-0.95in}&&   
{\cal N}_4 \, \,  = \,  \, \, 
(x+2) \cdot \, F(x)^3 \cdot \, {{d {\cal N}_3} \over {dx}} \,  \,   \, 
-F(x)^2 \cdot \,  
\Bigl(F(x) \, + 3 \cdot \, (x+2) \cdot \, F'(x) \Bigr) \cdot {\cal N}_3.
\end{eqnarray}

If ${\cal N}_1 = 0$ necessarily  ${\cal N}_2 = 0$ and 
${\cal N}_3 = 0$. Similarly, if ${\cal N}_3 = 0$
necessarily  ${\cal N}_4 = 0$ (but not conversely).

\vskip .1cm 

All these  non-linear differential equations 
(\ref{extremelyequ}), (\ref{otherequ}), (\ref{anotherequ}), 
(\ref{lastequ}) share the same solution (\ref{extremely}), 
and have movable singularities.  

\vskip .1cm 

In contrast with linear differential equations
one can have, for a given series,
 many non-linear differential equations of the same order,
for instance ${\cal N}_2 = 0$ and ${\cal N}_3 = 0$.
For linear differential equations, the (unique) minimal 
order linear ODE requires (paradoxically~\cite{Khi6}) 
many more coefficients to be obtained from 
the so-called ``guessing procedures'' than higher order ODEs.
When we use our (non-linear guessing) program it is not clear if 
the minimal order non-linear ODEs are the easiest to be obtained, 
{\em requiring the minimal number of coefficients to be ``guessed''}. 
Furthermore, higher order ODEs could be of smaller ``size''. 

\vskip .1cm 

\subsubsection{Running the program: a third simple example of a ratio of holonomic functions. \\}
\label{running3}

The  non-linear differential equation (\ref{suchthat}) 
corresponding to the simple example given in~\cite{Selected} 
of the ratio of two hypergeometric functions 
\begin{eqnarray}
\label{reduc7A}
\hspace{-0.75in}&& \quad \quad \quad \quad \quad \quad
R(x) \, \, = \, \,  \,  \, \, 
{{ _2F_1\Bigl(\bigl[{{1} \over {3}}, \,{{1} \over {3}}\bigr], 
\, [1], \, \, 27 \, x \Bigr) 
} \over {
 _2F_1\Bigl(\bigl[{{1} \over {2}}, \,{{1} \over {2}}\bigr],
 \, [1], \, \, 16 \, x \Bigr) }},
\end{eqnarray}
is obtained with our program in about 5 seconds. Note that this non-linear 
differential equation (see equation (D.2) in Appendix D of~\cite{Selected})
can be rewritten in a form that makes the emergence of a 
{\em Schwarzian derivative} crystal clear: 
\begin{eqnarray}
\label{reduc7A}
\hspace{-0.95in}&&  \, 
2\,{x}^{2} \cdot \, (1 \, -27\,x)^{2} \cdot \, (1\, -16\,x)^{2} \cdot \, {\, R_{1}}^{2}
 \,  \cdot \, 
\Bigl({\frac {\, R_{3}}{\, R_{1}}} \, -{{3} \over {2}} \,{\frac {{\, R_{2}}^{2}}{{\, R_{1}}^{2}}}\Bigr) 
\nonumber \\
\hspace{-0.95in}&& \quad  \, 
+2\,{x}^{2}
 \cdot \, (1 \, -16\,x)  \cdot \, (1 \, -27\,x)  \cdot \,(1\, +72\,x)   \cdot \, R \cdot \, R_{2}  \cdot \, 
\, \Bigl(3\,{\frac {\, R_{1}}{R}} \, -{\frac {\, R_{3}}{\, R_{2}}}\Bigr)
\nonumber \\
\hspace{-0.95in}&& \quad  \,  
\,  + \, (1\, -16\,x) 
  \cdot \, (1944\,{x}^{3}-1569\,{x}^{2}+58\,x-1) \cdot \,  {\, R_{1}}^{2}
 \\
\hspace{-0.95in}&& \quad  \, 
\, -2 \cdot \,x \cdot \, (93312\,{x}^{3}-168\,{x}^{2}-297\,x+4) \cdot \, R \cdot \, R_{2}
\nonumber \\
\hspace{-0.95in}&& \quad  \, 
\, -2 \cdot  \, (29376\,{x}^{3}+5580\,{x}^{2}-221\,x+1) \cdot \,  R \cdot \, R_{1} 
\,\, \, + \,  44\,{x}^{2}-432\,x+1) \cdot \,  {R}^{2} 
 \,\,  \, = \, \, \, 0.
\nonumber 
\end{eqnarray}

\subsubsection{Running the program: a fourth simple example of a ratio of holonomic functions. \\}
\label{running4}

The non-linear differential 
equation of the form (\ref{suchthat}) for 
a ratio of ${}_4F_3$ and ${}_2F_1$ hypergeometric functions
such as 
\begin{eqnarray}
\label{CalabiYau}
\hspace{-0.75in}&& \quad \quad \quad \quad \quad \quad
R(x) \, \, = \, \,  \,  \, \, 
{{ _4F_3\Bigl(\bigl[{{1} \over {2}}, \,{{1} \over {2}},
 \,{{1} \over {2}}, \,{{1} \over {2}}\bigl], 
\, [1, \, 1, \, 1], \, \, 256 \, x \Bigr) 
} \over {
 _2F_1\Bigl(\bigl[{{1} \over {2}}, \,{{1} \over {2}}\bigl],
 \, [1], \, \, 16 \, x \Bigr) }}, 
\end{eqnarray}
is obtained in 2834 seconds by our program. The  
non-linear differential equation (\ref{suchthat}) is of 
order five, and of degree $12$ in $x$ 
and is a (quadratic) homogeneous polynomial in $R(x)$, 
and its derivatives $R'(x), \ldots, R^{(5)}$.

\vskip .1cm

\subsubsection{Running the program: the composition of holonomic functions. \\}
\label{compoholo}

The composition of two ${}_2F_1$ hypergeometric functions, 
given in Section (\ref{composidiag}), 
is the solution of a non-linear differential equation 
(\ref{suchthat}) of order four, and of degree $32$ in $x$ 
and of degree at most $m= 11$ in $F(x)$, and its derivatives
$F'(x), \ldots, F^{(4)}$. With $m=  11$, there are 
${m+4\choose 4}  =  1365$ monomials 
$F(x)^{c_i(0)}  F'(x)^{c_i(1)} \cdots  F^{(4)}(x)^{c_i(4)}$
of degree  exactly 
$ d_c = c_i(0)+c_i(1)+c_i(2)+c_i(3)+c_i(4)  = 11$,
 in  $ F(x)$, and its derivatives
$F'(x), \ldots, F^{(4)}$. There are ${10 +4\choose 4}  =  1001$ 
monomials of degree $ d_c$  exactly equal to 
$10$, $ {9 +4\choose 4} =   715$ 
monomials of degree $ d_c$  exactly equal $ 9$, etc., 
and therefore there are 
$  {11 +4\choose 4}  +  {10 +4\choose 4}  + {9 +4\choose 4} 
 + \cdots = {11 +5\choose 5}  =  4368$
monomials of degree less than or equal to $ m= 11$. 

Since the degree in $x$ is $32$
the composition of two ${}_2F_1$ hypergeometric functions given in 
Section (\ref{composidiag}) would require a series expansion with 
$4368 \cdot 33  = 144144$ coefficients to find 
the non-linear differential equation (\ref{suchthat}).
Another program specially built to detect 
non-linear differential equations (\ref{suchthat}) 
that are homogeneous polynomials in $ F(x)$, and its derivatives
$F'(x), \ldots, F^{(4)}$ would require 
$ 1365 \cdot 33  =  45045$ coefficients to find 
the non-linear differential equation.

\vskip .1cm

In general finding the non-linear differential equation (\ref{suchthat})
of order $ k$, of degree $ d$ in $ x$, of degree at most $ m$ in 
$ F(x)$, and its derivatives $ F'(x), \ldots, F^{(k)}$ would require
\begin{eqnarray}
\label{binomialformula}
\hspace{-0.75in}&& \quad \quad \quad \quad \quad \quad
 (d+1) \cdot  \, {m  \,+k \, +1 \choose  k  \,+ \, 1} 
\end{eqnarray}
coefficients to be found with our program.

\vskip .1cm

Finding the  non-linear differential equation (\ref{AH2mod5bis}) 
solving the example of an algebraic function of holonomic functions 
(\ref{rootsser}), displayed in Section (\ref{moreinvolvedalg}), would 
require a really  large number of series coefficients.
     
\vskip .1cm
  
\vskip .1cm

\section{Application to the Ising series.}
\label{Appli}

We applied the algorithm described above to the $5043$ known 
initial terms of the square-lattice Ising susceptibility series, 
using both the high- and low-temperature series. As the generating 
functions for the two Ising series are believed to have a natural 
boundary, they cannot be algebraic or D-finite. This allows us 
to eliminate any forms $(m,k,d)$ with $k=0$ or $m=1$. This restriction 
leaves  $133$ possible polynomials $P(x, \, y_0,\,  \ldots, \, y_k)$. 
     
 \vskip .1cm
  
Using approximately $25$ hours of computation time on a machine 
with $256$ gigabytes of RAM, we have determined that neither 
the high- nor low-temperature square-lattice Ising susceptibility 
series satisfy an algebraic differential 
equation corresponding to any possible predicted form, as described above, 
with $U + k + T \leq 5043$. We repeated the computation with the series 
obtained by shifting the coefficients of the two given series one, two, 
and three positions (i.e., dividing by $x$, $x^2$, and $x^3$), because 
the number of terms needed to guess a functional equation is very 
sensitive to the positioning of the first non-zero term of the series.
     
 \vskip .1cm
  
This provides a collection of negative results. For example, if the 
susceptibility series are differentially algebraic with algebraic and 
differential order $3$ (i.e. $m=3$ and $k=3$), then it must be that 
the polynomial coefficients have degree strictly greater than 
$142$, otherwise our algorithm would have found the algebraic 
differential equations that they satisfy.
       
 \vskip .1cm
            
\section{Conclusion: To be or not to be differentially algebraic.}
\label{conclusion}

The preceding examples allow us to better understand the class 
of power series with integer coefficients, no Hadamard gaps, 
and non-zero radius of convergence,  
that reduce to algebraic functions modulo $p^r$. This is an 
extremely large class of functions probably with a closure 
property with respect to the composition of functions. 
The question whether this class of functions reduces 
to differentially algebraic functions remains a (difficult) open 
question. It contains the class of functions generated by the 
composition of algebraic functions of diagonals of rational functions, 
these functions always being differentially algebraic. 
     
 \vskip .1cm
  
In summary, the question that is raised is if the full 
susceptibility of the Ising model (and more generally other non-holonomic 
functions emerging in lattice statistical mechanics,
 or enumerative combinatorics~\cite{Selected}) is 
{\em differentially algebraic or is differentially transcendental}\footnote{A 
function that is not differentially algebraic is called differentially 
transcendental.}. If the  full susceptibility of the Ising model is 
differentially transcendental, the reduction modulo $ 2^r$ 
of the  full susceptibility series to algebraic 
functions seen in~\cite{Auto}, should just be seen as a consequence 
of a {\em specific character of powers of two} for the $\chi^{(n)}$, 
each of these $n$-fold integrals being series with integer 
coefficients up to an overall $2^r$ factor. This is already 
a remarkable property (see~\cite{Selected}). 
A differentially transcendental full susceptibility has, of course, 
no reason to reduce to algebraic functions modulo other primes, 
or powers of primes. However, one would now have to 
understand how such a {\em differentially transcendental} 
full susceptibility series could 
actually correspond to ``{\em algorithmic integrability}'', 
these series being obtained from an iterative algorithm having  
{\em polynomial growth}!

That is to say, the susceptibility series coefficients $ c_n$ are 
obtained from an algorithm with {\em polynomial growth} which 
is a consequence of a quadratic recurrence relation~\cite{Perk,Perk2,Perk3} 
on the two-point correlation functions $ C(M,  N)$
(recursion on the two integers $ M$ and $ N$). It is also tempting 
to speculate on the existence of a non-linear recursion, or more 
generally a polynomial relation on a fixed finite number $ p+1$ 
of coefficients, of the type: 
 \begin{eqnarray}
 \label{recnonlin}
 \hspace{-0.75in}&& \quad \quad \quad \quad \quad \quad
  P(n, \, c_n, \,  c_{n-1}, \,  c_{n-2}, \, \cdots \, c_{n-p})
  \, \, = \, \, \, 0. 
 \end{eqnarray}
Such a non-linear recursion should be compared with the non-linear 
ODE (\ref{glo2}) defining a differentially algebraic function. 
One can always convert a linear recurrence (with a finite number 
of terms) into a linear differential equation:
in the linear case, there is a ``duality'' 
between these two concepts. In contrast the correspondence between 
non-linear recursions and non-linear ODEs is more involved. Non-linear 
ODEs yield non-linear recursions, but these recursions are, not, 
in general, relations on a {\em fixed} number of terms like 
(\ref{recnonlin}). For instance, Tutte's non-linear ODE 
(\ref{Tutteq}) yields a quadratic recursion (\ref{Tutterecq}) 
where the number of terms involved grows 
at each step. Conversely, the existence of a non-linear 
recursion of the form (\ref{recnonlin}) is not a guarantee 
that the corresponding power series is differentially 
algebraic. One can imagine the existence of a non-linear recursion
of the form (\ref{recnonlin}) for the susceptibility series that could 
be obtained quite reasonably, while at the same time 
the non-linear ODE (\ref{1dalg})  proving 
the differentially algebraic character of the  susceptibility series
being extremely difficult to identify (if it exists!). 

It now becomes useful to build more efficient programs 
(see Section \ref{nodiffalg}) 
to see when such very long series, with more than 5043 coefficients, 
are differentially algebraic, or correspond to a non-linear 
recursion such as (\ref{recnonlin}). Recalling the previous examples 
(\ref{rootsser}), (\ref{seriescompoH1H2}), for which the polynomial 
differential equations were, {\em even for these extremely simple examples}, 
quite large polynomials, sometimes of quite high degrees,  
one can imagine, if the full susceptibility of the Ising model
turns out to be differentially algebraic, that the corresponding  
polynomial differential equation might be quite a large 
expression (recall the $144144$ coefficients required 
to find the non-linear differential equation (\ref{suchthat}) for 
a simple composition of two ${}_2F_1$ hypergeometric function). In 
the most pessimistic scenario such programs (that could be programs 
modulo primes)
should, at least,  give some lower bounds for the various degrees 
of the polynomial. Actually we already have a first modulo prime 
version of our program. We checked with 5043 terms modulo 3 
and, similarly, did not find any differentially algebraic 
equation modulo 3. Determining if the square-lattice Ising 
susceptibility is a differentially algebraic function is 
clearly a serious challenge. In contrast, non-holonomic 
generating functions of examples from
{\em enumerative combinatorics}~\cite{Bernardi2,Bousquetcat,Courtiel,Tutte1,Tutte2,Tutte3}  
may be a much simpler testing ground for such programs.

\vskip .1cm 

\vskip .2cm

\vskip .2cm

{\bf Acknowledgments:} JMM would like to thank J-P. Allouche, C. Banderier,
G. Christol, S. Fischler, T. Rivoal, J. Rocques and J. Sauloy J-A. Weil 
for fruitful discussions on differentially algebraic functions.  
JMM would like to thank D. Bertrand for providing a proof by Ehud Hrushovski,
on the closure by composition of differentially algebraic functions. He also 
thanks  C. Koutschan and M. Mezzarobba for discussions on zeros of D-finite
functions. This work has been performed without
 any support of the ANR, the ERC, the MAE or any PES of the CNRS. 
AJG and IJ would like to thank the Australian Research Council 
for its support through grant DP140101110.  IJ was
supported by an award under the Merit Allocation Scheme of the NCI
National facility at the ANU, where the bulk of the large scale
numerical computations were performed.   
JP would like to thank the Centre of Excellence 
for Mathematics and Statistics of Complex Systems for their 
hospitality in June 2015, where this research was partially 
conducted. JP was supported by the National Science Foundation 
and the Australian Academy of Science under award \#1514825.

\vskip .1cm
\vskip .1cm

\vskip .2cm

\appendix

\section{Differentially Finite and Differentially Algebraic Series}
\label{appendix:defs}
Let $\mathbb K$ be a field of characteristic zero (usually 
$\mathbb{K}=\mathbb{Q}$ 
or $\mathbb{K}=\mathbb{C}$). 

\subsection{D-finite or holonomic functions} 
\label{G1}

A function or formal power series $f(x)  \in \mathbb{K}[[x]]$ 
satisfying a linear differential equation
\begin{eqnarray}
\label{glo1}
\hspace{-0.95in}&&  \, \,\, \, \quad    \quad    \quad     
a_n(x)\cdot \, f^{(n)}(x)\,\,  + a_{n-1}(x) \cdot \,f^{(n-1)}(x)
\,\,  + \ldots \, \, + a_0(x)\cdot \,f(x) 
\,\,\,\,   = \,\, \, \,  0, 
\nonumber 
\end{eqnarray}
with polynomials $ a_i(x) \in \mathbb{K}[[x]]$ is called 
{\it differentiably finite} or D-finite or {\it holonomic}. The 
set of all D-finite power series is closed under addition, 
multiplication, {\em right composition with algebraic functions}, 
Laplace Borel transform or inverse Borel transform.
Lipshitz~\cite{L88} proved that the diagonal of a D-finite power 
series in several variables is D-finite, from which follows that 
the Hadamard product of a rational power series with 
a D-finite power series is D-finite.

A function that can be expressed as the diagonal of a rational 
function is D-finite. The converse is false: for instance, a 
D-finite function with an irregular singularity, such as 
$\, \exp(-1/(1-x))$, is a counter example.

A function that is algebraic {\em modulo} a prime $p$ 
or a power of a prime $p^r$ may be D-finite, but need 
not be. It also may be expressible as the diagonal 
of a rational function.

The ratio of two (non-algebraic) holonomic functions 
is generically {\em not} holonomic: it is differentially 
algebraic. The composition of two non-algebraic 
holonomic functions is {\em not} holonomic, though 
it is differentially algebraic. 

\vskip .1cm 

\subsection{Differentially algebraic functions} 
\label{G2}

Let $P$ be a non-zero polynomial in $n+2$ variables with 
coefficients in a field of characteristic zero 
$\mathbb{K}$, such that 
\begin{eqnarray}
\label{glo2}
\hspace{-0.95in}&&  \, \, \quad    \quad   \quad  
 \quad      \quad   \quad    \quad    
P(x, \, f(x), \,\, f'(x), \,\, \ldots, \, \, f^{(n)}(x)) 
\,  \,\, = \,  \,\, \, \, 0.
\end{eqnarray}
This is called an {\em algebraic differential equation}, and 
the function $ \, f(x)  \, \in \, \mathbb{K}[[x]]$ satisfying the 
algebraic differential equation  is called 
{\it differentially algebraic} or D-algebraic.
Lipshitz and Rubel~\cite{Lipshitz} proved that the set of all D-algebraic 
power series is  closed under addition, subtraction, multiplication 
and division (suitably restricted so that the expressions are sensible) 
{\em and composition of functions}. This set 
{\em is not, however, closed under Hadamard products}, 
and it is not closed under the 
{\em Laplace Borel transform} or the {\em inverse Borel transform}.
One simple example is, for instance, the Borel transform
of the generating function of the Bell numbers. 

\vskip .1cm

Formal series solutions of differentially algebraic
 equations are necessarily {\em Gevrey series}~\cite{Maillet} 
(see also the first footnote in Yves Andr\'e survey~\cite{Andre})
 i.e. the growth of the coefficients is bounded
 by  $\, C^n \cdot (n!)^s$. Conversely, if a series is 
``too divergent'' (not Gevrey) it 
{\em cannot be differentially algebraic},
it is differentially transcendent.

\vskip .1cm

\subsection{Diagonals of rational functions} 
\label{G4}
A power series in $ k$ variables is written 
\begin{eqnarray}
\label{glo3}
\hspace{-0.95in}&&  \, \, \quad   \quad    \quad    \quad   
  F(x_1,\, x_2, \, \cdots, \, x_k)
\,\, = \,\,\,\, \sum_{m_1} \sum_{m_2}\, \,\cdots 
\,\,
\sum_{m_k} \,\, a_{m_1,m_2, \cdots, m_k} 
\cdot \, x_1^{m_1}\, x_1^{m_2} \cdots \, x_k^{m_k}.
\nonumber 
\end{eqnarray}
Its {\em diagonal} is the one-variable series 
\begin{eqnarray}
\label{glo4}
\hspace{-0.95in}&&  \, \, \quad    \quad   \quad      \quad   
{\rm Diag}\left (F(x_1,\, x_2,\, \cdots, \, x_k)\right )
\,\, = \,\,\, \sum_m \,a_{m,m, \cdots, m}\,\cdot \, x^{m}.
\nonumber 
\end{eqnarray}
 Now if 
$ F(x_1,x_2, \cdots, x_k) =  \, P(x_1, x_2, \cdots, x_k)/Q(x_1, x _2, \cdots, x_k)$ 
where $ P$ and $ Q$ are polynomials with rational coefficients, and with 
$ Q(0,0, \cdots, 0) \ne 0$, then $ F(x_1, x_2, \cdots, x_k)$ is 
a rational function, and 
$ {\rm Diag}\left (F(x_1,\, x_2,\, \cdots, \, x_k)\right )$ 
is referred to as {\em the diagonal of a rational function.}

\vskip .1cm

 \subsection{Global boundedness} 
 \label{G5}

 A power series $\sum a_nx^n$ with rational coefficients $a_n$ 
and {\em non-zero radius of convergence} 
 is said to be {\em globally bounded} if there exists an integer $N$ 
such that the series  can be recast as one with integer coefficients 
by a rescaling $x \to N  x.$

Diagonals of rational functions, or diagonals of algebraic functions, 
are globally bounded series. Algebraic functions, being diagonals of 
rational functions~\cite{P21}, are globally bounded series. 

\subsection{Some relevant  theorems}
\begin{itemize}
\item
Eisenstein's Theorem~\cite{PS2} Chapter 3, 139. If a power 
series with rational coefficients 
represents an algebraic function 
it is globally bounded: it can be recast 
into a series which has integer coefficients.

\item
\cite{PS2} Chapter 3, 163. If a power series with integer coefficients 
represents a rational function, then, after a certain number of terms, 
the coefficients modulo $ m$ ($m$ arbitrary), form a {\em periodic} 
sequence (this result cannot be extended to algebraic functions).

\item
\cite{PS2} Chapter 3, 164. The algebraic function $1/{\sqrt{1 -4 x}}$ 
on expansion gives a power series with integer coefficients. These 
coefficients are for no odd prime number $p$ periodic modulo $p$.
\item
\cite{PS2} Chapter 3, 165. The radius of convergence of a non-terminating 
power series with integer coefficients is at most $1$.

\end{itemize}

\section{Solutions of Tutte's differentially algebraic equation 
(\ref{Tutteq}) modulo primes, for miscellaneous values of $ q$.}
\label{appendix1}

Let us consider the series (\ref{firstterms}) divided by $\, q \, (q-1)$:
\begin{eqnarray}
\label{firsttermsnormalised}
\hspace{-0.95in}&&  \quad \quad  \quad \quad \,  \, \, \,
{{H(x)} \over { q \, (q-1)}} \, \, \, = \, \, \, \,  \, 
  {x}^{2} \, \, \,  \, + \, \,  ( q-2 ) \cdot \, 
\sum_{n=3}^{\infty} \, \, P_n(q) \cdot \,  x^n. 
\end{eqnarray}

$\bullet$ The reduction modulo $3$ of this (normalised) 
series (\ref{firsttermsnormalised})  for $\, q=7$  
is an algebraic function:
\begin{eqnarray}
\label{q7ser11}
\hspace{-0.95in}&&  \, \, \, \, \quad \quad \quad \quad \quad
 F(x)^{3} \, 
\, +2\, {x}^{2}\cdot \,   F(x) \,\, \, 
+x^4 \cdot \, (1\, +2 \, x \, + x^2 \,+ x^5) 
\, \,\, \, = \, \, \, \,  \, 0.
\end{eqnarray}
The reduction modulo $ 7$ of the $ q=7$ series 
(\ref{firsttermsnormalised}) 
is the solution (modulo $ 7$) of a polynomial equation of degree 
$ 15$ in $ F(x)$, and degree $27 $ in $ x$.
Modulo $ 11$ the $ q=7$ series (\ref{firsttermsnormalised}) 
is the solution of a polynomial equation of degree 
$ 11$ in $ F(x)$, and degree $19 $ in $ x$.
Modulo $ 13$ the $ q=7$ series (\ref{firsttermsnormalised}) 
is  a solution of a polynomial equation of degree 
$ 15$ in $ F(x)$, and degree $ 27 $ in $ x$.

$\bullet$ The reduction modulo $ 5$ of the $ q=11$ 
series (\ref{firsttermsnormalised})
satisfies (modulo $ 5$)
\begin{eqnarray}
\label{q11ser5}
\hspace{-0.95in}&& \quad   
({x}^{25}+4)  \cdot \, F(x)^{4} \, \,  \, 
+x \cdot \, (2\,{x}^{25}+3)  \cdot \, F(x)^{3} 
\nonumber \\ 
\hspace{-0.95in}&&  \, \,    \quad  \quad    
\, \, 
\, +{x}^{2} \cdot \, (4\,{x}^{25} +1)  \cdot \, F(x)^{2} \, \, 
+{x}^{3} \cdot \, (3\,{x}^{25} +2) \cdot \, F(x) 
 \\ 
\hspace{-0.95in}&&  \, \,    \quad \quad      \, \, 
 +3\,{x}^{5} +{x}^{6}+4\,{x}^{7} +3\,{x}^{8} +2\,{x}^{9} +{x}^{10} 
+3\,{x}^{11}  +3\,{x}^{12}  +4\,{x}^{13} +{x}^{14}+4\,{x}^{15}
\nonumber \\ 
\hspace{-0.95in}&&  \, \,    \quad  \quad     \, \, 
+2\,{x}^{17} +3\,{x}^{18}  +4\,{x}^{19} +2\,{x}^{20} +4\,{x}^{21}
+3\,{x}^{23} +2\,{x}^{24} +2\,{x}^{25} +4\,{x}^{26} +3\,{x}^{27} 
\nonumber \\ 
\hspace{-0.95in}&&  \, \,   \quad  \quad    \, \,
 +3\,{x}^{28} +2\,{x}^{30} +4\,{x}^{32} +3\,{x}^{33}
+{x}^{34} 
\,   +{x}^{35} +3\,{x}^{36}+4\,{x}^{37}+{x}^{38}+{x}^{40}
\,\, = \, \,\, \, 0.
\nonumber 
\end{eqnarray}
The $ q=11$ series (\ref{firsttermsnormalised})
satisfies modulo $7$
\begin{eqnarray}
\label{q11ser7}
\hspace{-0.95in}&&  \, \, \quad  \,  \quad
F(x)^{4} \, \,  \, 
+ \, (5\,x+1)  \cdot \, F(x)^{3} \,\, \, 
+x \cdot \, (6\,{x}^{2} +5\,x +2)  \cdot \, F(x)^{2}
\nonumber \\ 
\hspace{-0.95in}&&  \, \,  \quad 
 \quad \quad \quad
\, +{x}^{2} \cdot \, ({x}^{2}+2\,x+6) \cdot \, F(x)\,
\,\, \,    +{x}^{4}  \cdot \, (1\,+5 \, x\, +2 \, x^2 ) 
\,  \,\, = \,\, \, \,\, 0.
\end{eqnarray}

$\bullet$ For the $ q=-1$ series, 
modulo $ 5$ the series (\ref{firsttermsnormalised}) 
is an algebraic function
\begin{eqnarray}
\label{qm1ser5}
\hspace{-0.95in}&&  \, \, \quad \quad \quad  \quad 
2\,  x^3 \cdot \, (2 \, +\, x \, +\, x^2 )
 \,  \, + x \cdot \, F(x)\, \, \,  + \, F(x)^{2}
 \, \,  \, = \,\,  \,  \, 0.
\end{eqnarray}
Modulo $ 7$ the $ q=-1$ series is also an algebraic function
\begin{eqnarray}
\label{qm1ser7}
\hspace{-0.95in}&&  \, \, \quad \quad\quad \quad
 F(x)^{5}\,
\, + \, (6\,{x}^{2}+x+3)  \cdot \,  F(x)^{4} \, \,
+ \, (2\,{x}^{4}+5\,{x}^{2}+6\,x+1)  \cdot \, F(x)^{3}
\nonumber \\ 
\hspace{-0.95in}&&  \, \, \quad   \quad \quad \quad  \quad \quad
\, +x \cdot \, (6\,{x}^{4}+5\,{x}^{3}+4\,{x}^{2}+x+3)
  \cdot \, F(x)^{2} 
\nonumber \\ 
\hspace{-0.95in}&&  \, \, \quad   \quad \quad \quad  \quad \quad
\, 
+{x}^{2} \cdot \, ({x}^{5} +6\,{x}^{4} +5\,{x}^{3} +5\,{x}^{2}+2)
 \cdot \, F(x)
\nonumber \\ 
\hspace{-0.95in}&&  \, \, \quad   \quad  \quad \quad \quad \quad
\,   +x^4 \cdot \, 
(5 \, +3 \, x \, +3 \, x^2 \, +x^3 \, +2 \, x^4 +4 \, x^5) 
\,  \, \, = \, \,  \,  \, 0.
\end{eqnarray}
Modulo $ 11$ the $ q=-1$ series 
is the solution of a polynomial equation of degree 
$ 7$ in $ F(x)$, and degree $13$ in $ x.$  
Modulo $ 13$ the $ q=-1$ series 
is the solution of a polynomial equation of degree 
$ 6$ in $ F(x)$, and degree $10$ in $ x$.
Modulo $ 17$ the $ q=-1$ series 
is the solution of a polynomial equation of degree 
$ 9$ in $ F(x)$, and degree $15$ in $ x$.

$\bullet$ For the $ q=  1/2$  series (\ref{firsttermsnormalised}) 
(when it works\footnote{These 
calculations cannot be performed modulo any integer.}), we find,
modulo $ 5,$ the algebraic function
\begin{eqnarray}
\label{q1over2ser5}
\hspace{-0.95in}&&   \, \quad \quad \quad
 \, F(x)^{2}
\,\, 
+ \left( x+1 \right) \cdot \, \left( 2\,x+1 \right) \cdot \, F(x)
\, \,\, +{x}^{2}  \cdot \, (4 \, +x \, +x^2)
\,\,\,\,   = \,\, \, \, 0, 
\end{eqnarray}
and modulo $ 7$
\begin{eqnarray}
\label{q1over2ser7}
\hspace{-0.95in}&&  \, \, \quad  \quad  \,  \, 
 F(x)^{4} \, \,+ \, (5\,x+1)  \cdot \, F(x)^{3} \,\, 
+x \cdot \, (6\,{x}^{2}+5\,x+2)  \cdot \, F(x)^{2}
\nonumber \\ 
\hspace{-0.95in}&&  \, \, \,  \,  \, 
 \quad  \quad  \quad \quad  
\, +{x}^{2} \cdot \, ({x}^{2}+2\,x+6) \cdot \, F(x) \, \,
\,  +x^4 \cdot \, (1 \, +5 \, x\, +2 \, x^2) 
\, \, \,\,  = \, \,  \,  \, \, 0.
\end{eqnarray}
Modulo $ 11$ the $ q= 1/2$ series 
is the solution of a polynomial equation of degree 
$ 7$ in $ F(x)$, and degree $13$ in $ x$.
Modulo $ 13$ the $ q=1/2$ series 
is the solution of a polynomial equation of degree 
$ 15$ in $ F(x)$, and degree $27$ in $ x$.
Modulo $ 17$ the $ q=1/2$ series 
is the solution of a polynomial equation of degree 
$ 9$ in $ F(x)$ and degree $ 15$ in $ x$.
Modulo $ 23$ the $ q=1/2$ series 
is the solution of a polynomial equation of degree 
$ 15$ in $ F(x)$, and degree $ 25$ in $ x$.

\vskip .1cm

\section{A simple generalisation of Tutte's differentially 
algebraic equation (\ref{Tutteq}).}
\label{appendix2}

The {\em two-parameter differentially algebraic equation}
(the two parameters are the two integers $ M$ and $ N$):
\begin{eqnarray}
\label{TutteMoverN}
\hspace{-0.98in}&&   
-2 \, M^2 \cdot \, (M\, -N)  \cdot \, x \,  \, \,  \,  \, 
+ \, \Bigl(M \cdot \, x \, + 10 \, H(x) \,
 -6 \, x \, { {d H(x)} \over {dx}} \Bigr)
 \cdot \,   { {d^2 H(x)} \over {dx}^2} 
 \nonumber   \\
\hspace{-0.95in}&&   \quad       \quad    
-\, M \cdot \, (M\, -4 \, N)  \cdot \, \Bigl( 20 \, H(x) \,
 - \,18 \, x \, { {d H(x)} \over {dx}} \,
 + 9 \, x^2 \, { {d^2 H(x)} \over {dx^2}} \Bigr)
\, \,\,\,  =  \, \, \,\, \, 0,  
\end{eqnarray}
has solution series of the form
\begin{eqnarray}
\label{TutteMoverNser}
\hspace{-0.95in}&&   
M \cdot \,  (M-N) \cdot \, {x}^{2} 
\,  \, +\, \sum_{n=3}^{\infty} \, M  \cdot \,(M-N)  \cdot \, 
(M-2\,N) \cdot \, P_{n}(M, \, N) \cdot \, x^n
\, \, \,\,  + \, \, \cdots 
\end{eqnarray}
where:
\begin{eqnarray} 
\hspace{-0.95in}&&    
P_3(M, N)  = \, 1,  \quad \,\,
P_4(M, N)  = \, \, 4\,M-9\,N, \quad \,\,
P_5(M, N)  = \, \, 8\,{M}^{2}-37\,MN+43\,{N}^{2},
\nonumber   \\
\hspace{-0.95in}&&   
P_6(M, N)  = \,\,176\,{M}^{3} -1245\,{M}^{2}N +2951\,M{N}^{2}
-2344\,{N}^{3},
\nonumber   \\
\hspace{-0.95in}&&    
P_7(M, N)  = \,\,1456\,{M}^{4}-13935\,{M}^{3}N
 +50273\,{M}^{2}{N}^{2}
-81036\,M{N}^{3}+49248\,{N}^{4},
\nonumber  
\end{eqnarray}
where the $ P_n$'s are homogeneous polynomials of $ M$ 
and $ N$ of degree $ n  -3$.
One can show that all these  two-parameter series with 
integer coefficients reduce to algebraic functions modulo 
every prime, or power of a prime. 

\vskip .1cm

\section{Miscellaneous differentially algebraic equations associated with 
simple quadratic recursions.}
\label{appendix3}

Recall that  $ H(x)  =  \sum  h_n  x^n$,
corresponding to  the series (\ref{firstterms}), 
can be obtained
from a simple 
{\em quadratic recurrence relation}:
\begin{eqnarray}
\label{Tutterecq}
\hspace{-0.95in}&&   \quad  \, \, \,\,
q \cdot \, (n+1)\cdot \,(n+2) \cdot \, h_{n+2} \, \,\, \, = \, \,  \, \,\,\,
q  \cdot \, (q\,-4)  \cdot \, (3\, n\, -1) \cdot \, (3\, n\, -2) \cdot \, h_{n+1} 
\nonumber  \\
\hspace{-0.95in}&&   \quad  \quad \quad \quad  \quad   \quad \quad  
\, + \, 2 \, \sum_{i=1}^{n} \, \,  i \cdot \, (i+1) \cdot \, 
(3\, n\, -3\, i\, +1) \cdot \,  h_{i+1} \, h_{n-i+2}, 
\end{eqnarray}
Such quadratic recurrence relations are clearly a very simple and 
efficient way to generate power series such that the constraint that 
each coefficient of the series is an integer is {\em guaranteed}. 

\subsection{A first example corresponding to divergent series.}
\label{appendix31}

Consider a slight modification
of the quadratic recurrence (\ref{Tutterecq}):
\begin{eqnarray}
\label{Tutterec4modif}
\hspace{-0.95in}&&   \quad  \quad  \,  \quad   \quad   
h_{n+2} \, \,\, \, = \, \,  \, \,\,
\,  \,  \, \sum_{i=1}^{n} \, \,  i \cdot \, (i+1) \cdot \, 
(3\, n\, -3\, i\, +1) \cdot \,  h_{i+1} \, h_{n-i+2}, 
\end{eqnarray}
Note that recurrences like (\ref{Tutterec4modif}) (not 
necessarily quadratic)
obviously yield series with {\em integer coefficients}.
For instance, with the initial conditions $ h_0 =  0$,  
$ h_1 =  0$, $ h_2 =  1$, the recurrence 
(\ref{Tutterec4modif}) yields the series:
\begin{eqnarray}
\hspace{-0.95in}&&  \, \, \quad  \quad  
x^2 \,  \,\, +2 \, \, x^3  \, \,\,+28 \,  \, x^4 \, \, \,
+824  \, \, x^5  \,\,\,  +38000 \,  \, x^6\,\,  \, 
+2424576  \, \, x^7 \,\, +200465344 \,  \, x^8 
\nonumber \\
\hspace{-0.95in}&&  \quad  \quad  \quad    \quad \quad \,\,
+20649137664  \, \, x^9 
\,\,+2581342891776 \, \, x^{10}
 \, \, \, \, \, +   \, \, \, \cdots 
\nonumber 
\end{eqnarray}
Note that this series is a {\em divergent series}.
Modulo $ 3$ this series reduces to an algebraic function: 
\begin{eqnarray}
\hspace{-0.95in}&&  \quad  \quad  \quad \quad  \quad  \quad   
2\,{x}^{7} \,\,  \,\, +x^5 \cdot \, F(x)  \,\, \,\, + F(x)^4 
\, \, \, \,= \, \,\, \, \, 0.
\end{eqnarray}
Modulo $ 5$ this series reduces to an algebraic function: 
\begin{eqnarray}
\hspace{-0.95in}&&  \quad  \quad  \quad \quad  \quad   \quad  
{x}^{10} \, \, \, \, +4\, x^8 \cdot \, F(x) \, \, \, \, 
+2 \, x^5 \cdot \, F(x)^3 \, \, +F(x)^6 \,
\, \, \,\, = \, \,\, \, \, 0.
\end{eqnarray}
Modulo $ 7$ this series reduces to an algebraic function
of degree $ 48$ in $ x$ and degree $ 31$ in $ F(x)$.

\vskip .1cm

\subsection{A second example corresponding to  a divergent series.}
\label{appendix32}

Now consider another modification
of the quadratic recurrence (\ref{Tutterecq}):
\begin{eqnarray}
\label{Tutterec4modif7}
\hspace{-0.95in}&&   \quad  \quad  \quad  
 \quad  \quad   \quad  \quad  
h_{n+2} \, \,\, \, = \, \,  \, \,\,
\,  \,  \, \sum_{i=1}^{n} \, \,   
i \,  \cdot \,  h_{i+1} \, h_{n-i+2}, 
\end{eqnarray}
yielding the divergent series
\begin{eqnarray}
\label{serTutterec4modif7}
\hspace{-0.95in}&&   \quad  \,  \, 
 {x}^{2} \, \, \, +{x}^{3} \, \, +3\,{x}^{4} \, \, +14\,{x}^{5} 
 \,\, +85\,{x}^{6} \, +621\,{x}^{7} \, \, +5236\,{x}^{8} 
\, \, +49680\,{x}^{9} \, \, +521721\,{x}^{10}
\nonumber \\ 
\hspace{-0.95in}&&   \quad  \quad   \quad  \quad  \quad 
 \,\,  +5994155\,{x}^{11} \, \, + \,  74701055 \,{x}^{12}
 \, \,\,  + \,  \, \cdots  
\end{eqnarray}
which is a solution of the differentially algebraic equation:
\begin{eqnarray}
\label{serTutterec4modif7}
\hspace{-0.95in}&&   \quad  \quad  \quad 
-x^3 \, \,  \, +x \cdot \, F(x) \, \, \, +F(x)^2 \, \, \,
-x \cdot \, F(x) \cdot \,{{ d F(x)} \over {d x}} 
\, \, \, = \,  \,\, \, \, 0. 
\end{eqnarray}
If one normalizes the divergent series (\ref{serTutterec4modif7}) 
by dividing through by $x^2$, it becomes
\begin{eqnarray}
\label{serTutterec4modif7norm}
\hspace{-0.95in}&&   \quad    \quad \, 
1 \, \,  \,+x\, +3\,{x}^{2}\, \, +14\,{x}^{3}\, \, +85\,{x}^{4}\,  \,
+621\,{x}^{5}\, \, +5236\,{x}^{6}\, \, +49680\,{x}^{7}\,  \,
+521721\,{x}^{8} 
\nonumber \\
\hspace{-0.95in}&&   \quad  \quad \quad \quad  \quad \, 
+5994155\,{x}^{9}  \, \, 
 + \,  74701055 \,{x}^{10} \, \,\,   \, + \, \, \cdots 
\end{eqnarray}
This normalized series is a solution of:
\begin{eqnarray}
\label{serTutterec4modif7norm}
\hspace{-0.95in}&&   \quad  \quad  \quad \quad  
x^2 \cdot \, y(x) \cdot \, {{ d y(x)} \over {d x}} \, \, \,
 +x \cdot \, y(x)^2 \,\, \, \,  -y(x) \,\, \, \,+1 
\, \,\,  \,= \, \,\, \, \, 0. 
\end{eqnarray}
The coefficients of (\ref{serTutterec4modif7}) (not the 
ones of (\ref{serTutterec4modif7norm}))
{\em are remarkably well approximated by}  $ c \cdot n!$ 
where $ c  \simeq  0.21795078$. 
This value is deduced from 9000 coefficients of the 
divergent series.  The  Borel transform 
 of the divergent series 
(\ref{serTutterec4modif7}) is very close to
\begin{eqnarray}
\label{Laplace}
\hspace{-0.95in}&&   \quad \quad \quad    \quad 
\simeq \,\,   \,\, 
{{ 0.21795078 } \over {1 \, -x}} 
\, \,\,   \,\,  
+ \, 0.65385 \cdot \, \ln(1\, -x). 
\end{eqnarray}
In general the Borel transform (or the inverse Borel transform) 
of a differentially algebraic function is {\em not}
a differentially algebraic function. However, here, the 
Borel transform  of the divergent series 
(\ref{serTutterec4modif7}) is so simple, we can imagine 
that it could be differentially algebraic. We have used our 
program on $2000$ and $ 3000$ coefficients of the Borel 
transform and inverse Borel transform of
(\ref{serTutterec4modif7}) and they do not seem to be
 differentially algebraic\footnote{Note that, similarly, using 
our program, the Borel and inverse Borel transform of Tutte's 
series (\ref{serq4}) which has a non-zero radius of convergence, 
do not seem to be differentially algebraic.}. 

Now consider perhaps the best-known
example of a {\em divergent series} namely 
$ \sum  n!  \cdot   x^n$,
which is a solution of the linear ODE\footnote{Which has a
solution like $ \exp(-1/x)/x$. The linear 
ODE (\ref{diveq}) can be obtained from 
the inverse Borel transform of $ y(x)  =   1/(1-x)$ using 
the $ {\tt invborel}((1-x)*y(x)-1, y(x), '{\rm diffeq}')$  
command in Maple. }:
\begin{eqnarray}
\label{diveq}
\hspace{-0.95in}&&   \quad  \quad  \quad \quad   \quad  \quad 
x^2 \cdot  {{ d y(x)} \over {d x}}
 \,  \,\, +(x\, -1) \cdot \, y(x) \,  \, \,\,
 +1 \,\, \, \,\, = \,\, \, \, 0. 
\end{eqnarray}
In order to be closer to 
(\ref{serTutterec4modif7norm}),  
we rewrite  this equation in a quadratic way:
\begin{eqnarray}
\label{diveqquadra}
\hspace{-0.95in}&&   \quad  \quad  \quad  \quad  \quad 
x^2 \cdot \, y(x)\cdot \, {{ d y(x)} \over {d x}}
 \,\,\, + \, (x\, -1)\cdot \, y(x)^2 \, \, \,
 +y(x) \, \,\, \, = \, \,\, \, 0. 
\end{eqnarray}
One remarks that this quadratic equation 
(\ref{diveqquadra}) {\em is extremely similar to
the differentially algebraic equation} 
(\ref{serTutterec4modif7norm}). 

Note that, again, the {\em divergent} series 
(\ref{serTutterec4modif7}) 
{\em reduces to algebraic functions modulo primes}.
Modulo $ 3$ it reduces to the algebraic function
defined by: 
\begin{eqnarray}
\label{divmod3}
\hspace{-0.95in}&&   \quad  \quad  \quad   \quad  \quad 
x^6 \, \,  \, +2\, x^4 \cdot \,  F(x) \, \,  \,  \,
 +x^3 \cdot \, F(x)^2 \,\,   \,   +F(x)^4
\, \,  \,\,   \, = \, \, \,  \, \, 0. 
\end{eqnarray}
Modulo $ 5$ it reduces to the algebraic function
defined by: 
\begin{eqnarray}
\label{divmod5}
\hspace{-0.95in}&&   \quad  \quad  \quad  \quad \, \, \, \,  
 F(x)^6 \,\, \, +x^3 \cdot \, F(x)^4 \, \,+2\,x^4 \cdot \, F(x)^3
\,\,\,   +x^5 \cdot \, (2\,x\,+1) \cdot \, F(x)^2
\nonumber \\ 
\hspace{-0.95in}&&   \quad  \quad  \quad \quad  \quad \quad  \quad 
\,+2\, x^6 \cdot \,(x+2) \cdot \, F(x) \,\,  \,
\, +\,x^8 \cdot \, (3 \, x\,+1)
\, \, \, \,= \, \,\,  \,\, 0. 
\end{eqnarray}
Modulo $ 7$ it reduces to the algebraic function
defined by: 
\begin{eqnarray}
\label{divmod7}
\hspace{-0.95in}&&   \quad  \, \,  \,\,    
F(x)^8 \, \,\, +x^3 \cdot \, F(x)^6 \, \,\,
+2\, x^4 \cdot \, F(x)^5 \,   \,  \, 
+3\, x^5 \cdot \,(x+2) \cdot \,F(x)^4 \, 
\nonumber \\ 
\hspace{-0.95in}&&   \quad   \,\,     \quad   \quad  \,
 \, +x^6 \cdot \,(4\,x \, +3) \cdot \,F(x)^3 \,
+x^7\cdot \, (4\, x^2 \, +x+1) \cdot \, F(x)^2 \,  \, 
\nonumber \\ 
\hspace{-0.95in}&&   \quad  \,\,    \quad   \quad 
 \,  \, 
+x^8 \cdot \, (5\,x^2\,+4\,x\,+6)\cdot \, F(x)\,
\, +x^{10} \cdot \,(1\,+x) \cdot \,(1\,+2 \,x) 
 \, \, \, = \, \,\,  \, \, 0. 
\end{eqnarray}

Note that the reduction modulo a prime of the divergent series 
$ \sum  n!  \cdot   x^n$, in contrast, always gives 
{\em polynomials}. For instance modulo $ 7$ it yields:
\begin{eqnarray}
\label{poldivmod7}
\hspace{-0.95in}&&   \quad  \quad  \quad  \quad  \quad  \quad 
1 \,\,\, +x \,\, +2\,{x}^{2} \,\, +6\,{x}^{3} \,\, +3\,{x}^{4}\,
 \, +{x}^{5} \,\, +6\,{x}^{6}.
\end{eqnarray}

\vskip .1cm 

\section{Algebraic functions of diagonals of rational functions}
\label{AFD}

It was noted in a previous paper (see equations (47) and (48),
in Section 5.2 of~\cite{Selected}),  that the two diagonals 
of rational functions $ H_1(x)$ and $ H_2(x)$ (\ref{twodiag})
reduce, modulo a prime, to simple algebraic 
functions.
For instance, modulo $ 7$, the two diagonals 
of rational functions (\ref{twodiag}) identify with the 
series expansion of the following algebraic functions:
\begin{eqnarray}
\label{twodiag1}
\hspace{-0.95in}&&   \quad   \quad  
 _2F_1\Bigl(\Bigl[{{1} \over {2}}, \, {{1} \over {2}}\Bigr], \, [1], 
\, 16 \cdot 20 \cdot \, x^2\Bigr) \, \,= \, \, \,
(1\,+3\,x^2\,+x^4\,+6\,x^6)^{-1/6}
 \nonumber \\ 
\hspace{-0.95in}&&   \quad   \quad    \quad   
\quad   \quad  \quad  \quad  \, = \, \, \,\,\,
1\,\,\,+3\,x^2\,\,+x^4\,+6\,x^6 \, + \,3\, x^{14} \,\, + \,2\, x^{16}
 \,\,\, + \, \, \cdots 
\end{eqnarray}
and
\begin{eqnarray}
\label{twodiag2}
\hspace{-0.95in}&&   \quad    \quad \quad   \quad  
 _2F_1\Bigl(\Bigl[{{1} \over {3}}, \, {{1} \over {3}}\Bigr], \, [1], 
\, 27 \cdot 20 \cdot \, x^2\Bigr) \, \,= \, \,\, 
\, (1\,+ 4 \, x^2\,+x^4)^{-1/6} 
 \nonumber \\ 
\hspace{-0.95in}&&   \quad  \quad   \quad 
 \quad   \quad   \quad  \quad  \quad  
\, = \, \, \,\, \, \,
1\,\,\,+4 \, x^2\,\,+x^4\,+4\, x^{14} \,+2\, x^{16} 
\,\, \,\,+ \,\, \cdots 
\end{eqnarray}
Denote by $ A_1$ and $ A_2$ the two algebraic functions
$ (1+3x^2+x^4+6x^6)^{-1/6}$ and $ (1+ 4  x^2+x^4)^{-1/6}$ respectively.
It is natural to expect that the reduction modulo $ 7$ 
of the non-holonomic series (\ref{rootsser}), namely 
\begin{eqnarray}
\label{reducmod7}
\hspace{-0.95in}&&   
1 \,+4\,x \,+3\,{x}^{2} \,+3\,{x}^{3}\,+{x}^{4}\, 
+2\,{x}^{5} \,+6\,{x}^{6} \,+4\,{x}^{7}
+5\,{x}^{9} \,+4\,{x}^{11}\, +5\,{x}^{13}\,
 \,\, + \,  \cdots 
\end{eqnarray}
is a solution  of (\ref{roots}) where $ H_1$ and $ H_2$ are 
replaced by the two algebraic functions $ A_1$ and $ A_2$, 
namely solution of 
\begin{eqnarray}
\label{rootsalg}
\hspace{-0.95in}&&   \quad  \quad \quad \quad \quad  \quad  \quad 
 z^2 \,\,  - \,\,2   \, A_1 \cdot \, z\,\,  +\,\, A_2 \, 
 \, \,\, = \, \, \,\, 0, 
\end{eqnarray}
where: 
\begin{eqnarray}
\label{A1A2alg}
\hspace{-0.95in}&&   \,\,
 (1\,+3\,x^2\,+x^4\,+6\,x^6) \cdot \, A_1^6 \, - \, 1 
\, = \,\,\,  \, 0, 
\quad \,  \,  \,  \, (1\,+ 4 \, x^2\,+x^4) \cdot \, A_2^6 \, - \, 1
\, \, = \,\,  \,\,\, 0.
\end{eqnarray}
Calculating  resultants between (\ref{rootsalg}) and the two polynomials 
(\ref{A1A2alg}), in order to eliminate $ A_1$ and $ A_2$, one gets, 
modulo $ 7$, a polynomial relation $\,  P_{60,  72}(x,  z)  =   0 \, $ 
of degree $ 60$ in $ x$ and $ 72$ in $ z$. As must be the case, this 
polynomial is, in fact, a function of $ x^2$. One can check that, 
modulo $ 7,$ the non-holonomic series (\ref{reducmod7})
is actually a solution of
$ P_{60,  72}(x,  z)  =   0 $ modulo $ 7$.

\vskip .1cm 

\section{A simple example of the composition of two holonomic 
functions modulo various primes.}
\label{subclosureapp}

Let us recall (\ref{compoH1H2}) the composition of the two 
 holonomic functions (that are diagonals of rational functions) 
(\ref{H1H2}). The corresponding series has been shown to reduce 
to algebraic functions modulo $ 5$. Of course there is nothing 
special about the prime $ 5$. For instance, modulo $ 3$ 
the series (\ref{seriescompoH1H2}) for function 
(\ref{compoH1H2}) becomes
\begin{eqnarray}
\label{v1mod3}
\hspace{-0.95in}&&   
1 \, +x \, +{x}^{2} \, +{x}^{3} \, +2\,{x}^{4} \, +2\,{x}^{5} \, +{x}^{6}
+2\,{x}^{7}+2\,{x}^{8} +{x}^{9}+2\,{x}^{12}+2\,{x}^{13}+2\,{x}^{14}
\nonumber \\ 
\hspace{-0.95in}&&   \quad \, 
+2\,{x}^{15} \,  +{x}^{18} \, +2\,{x}^{19} \, +2\,{x}^{20}+2\,{x}^{21}
+2\,{x}^{24}+2\,{x}^{25}+2\,{x}^{26} \, +{x}^{27}
\, \, + \, \, \, \cdots 
\end{eqnarray}
which is  an {\em algebraic function} given by the 
solution of the polynomial equation:
\begin{eqnarray}
\label{H2mod3}
\hspace{-0.95in}&&   \quad \quad \quad 
({x}^{2}+2\,x+2)  \cdot \, F(x)^{4} \, \, 
\, +2 \cdot \, (x+1) \cdot \, F(x)^2 \,\,  \, +2 \cdot \, (x+1)
\, \,\,  \, = \,\, \,  \, \,  0.
\end{eqnarray}

Modulo $ 7$ the series (\ref{seriescompoH1H2})
is, again, an algebraic function, given by the solution of 
 the polynomial equation:
\begin{eqnarray}
\label{v1mod7}
\hspace{-0.95in}&&     \, \, \, \,  \,  \quad  \quad  
p_{36} \cdot \,  F(x)^{36} \, 
\,\, + \, p_{30} \cdot  F(x)^{30} \, \,
\, \, + \, \, p_{24}  \cdot \,  F(x)^{24} \, \, \, \, 
+ \,  p_{12} \cdot  \,  (F(x)^{18}\, + F(x)^{12})
\nonumber \\ 
\hspace{-0.95in}&&   \quad  \quad  
\quad   \quad   \quad  \quad \quad 
 + \,  p_{6}
 \cdot \,   (F(x)^{6}\,  +1) 
\,  \, \, \,\, = \,\,  \, \, \,  0,
\end{eqnarray}
where:
\begin{eqnarray}
\label{v1mod7pol}
\hspace{-0.95in}&&    \quad   \quad\,
p_{36} \,\, = \, \,\,\, {x}^{18}\,+4\,{x}^{15}\,+4\,{x}^{14}\,
+2\,{x}^{13}+6\,{x}^{11}+6\,{x}^{10}+{x}^{9}+2\,{x}^{8}
+2\,{x}^{7}+3\,{x}^{6}
\nonumber \\ 
\hspace{-0.95in}&&   \quad  \quad \quad  \quad  \quad
+4\,{x}^{5}+2\,{x}^{4}+5\,{x}^{2}+2\,x \, +6,
\nonumber \\ 
\hspace{-0.95in}&&   \quad \quad \,
p_{30} \,\, = \, \,\,\,6\,{x}^{15}+6\,{x}^{14}+3\,{x}^{13}
+{x}^{12} +4\,{x}^{11}+4\,{x}^{10}+5\,{x}^{9}
+5\,{x}^{8}+{x}^{7}+2\,{x}^{6}
\nonumber \\ 
\hspace{-0.95in}&&   \quad  \quad \quad  \quad  \quad
+4\,{x}^{5}+2\,{x}^{4}+5\,{x}^{2}+2\,x\,+6, 
\nonumber \\ 
\hspace{-0.95in}&&   \quad \quad \,
p_{24} \,\, = \, \,\, \, 4\,{x}^{15} +4\,{x}^{14} 
+2\,{x}^{13} +{x}^{12} +{x}^{11} +{x}^{10} +4\,{x}^{9}
+6\,{x}^{8} +3\,{x}^{7} +4\,{x}^{6}
\nonumber \\ 
\hspace{-0.95in}&&   \quad  \quad \quad  \quad  \quad
+4\,{x}^{5} +2\,{x}^{4} +5\,{x}^{2} +2\,x\,+6, 
\nonumber \\ 
\hspace{-0.95in}&&   \quad \quad \,
p_{12} \,\, = \, \,\,\, 3\,{x}^{12} +6\,{x}^{11} 
+6\,{x}^{10} +{x}^{9} +2\,{x}^{8} +2\,{x}^{7}
+3\,{x}^{6}
\nonumber \\ 
\hspace{-0.95in}&&   \quad  \quad \quad  \quad  \quad
+4\,{x}^{5} +2\,{x}^{4} +5\,{x}^{2} +2\,x\, +6, 
\nonumber \\ 
\hspace{-0.95in}&&   \quad  \quad \,
p_{6} \,\, = \, \,\,\,
6\,{x}^{9}+4\,{x}^{8}+6\,{x}^{7}+4\,{x}^{5}+2\,{x}^{4}
+5\,{x}^{2}+2\,x\,+6.
\end{eqnarray}
Modulo $ 11$ the series (\ref{seriescompoH1H2})
is yet again, an algebraic function, given by the 
solution of the polynomial equation:
\begin{eqnarray}
\label{v1mod11}
\hspace{-0.95in}&&     \, \, \, \, 
 \quad  \quad \quad  \quad \quad   \quad  \quad    
\sum_{n=0}^{n=10} \, q_{n} \cdot \,  \Bigl(F(x)^{10} \Bigr)^n
\,  \, \, \, = \,\,  \, \, \,  0,
\end{eqnarray}
where $ q_{2} = q_{3}$, 
$ q_{0} = q_{1} $ and :
\begin{eqnarray}
\label{v1mod11pol}
\hspace{-0.95in}&&    \quad  
q_{10}\, \, \, = \,\,  \, \, 
{x}^{50} +9\,{x}^{45} +{x}^{44} +3\,{x}^{43} +5\,{x}^{42} 
+3\,{x}^{41} +10\,{x}^{40} +10\,{x}^{39} +2\,{x}^{37} +10\,{x}^{36}
\nonumber \\ 
\hspace{-0.95in}&&   \quad \quad \, 
+10\,{x}^{35} +3\,{x}^{34}
+10\,{x}^{32} +9\,{x}^{31} +3\,{x}^{30} +10\,{x}^{29} 
+7\,{x}^{28} +9\,{x}^{27} +8\,{x}^{26} 
\nonumber \\ 
\hspace{-0.95in}&&   \quad \quad \, 
+3\,{x}^{25} +4\,{x}^{24} +4\,{x}^{23}+5\,{x}^{21}
+6\,{x}^{20} +4\,{x}^{19} +8\,{x}^{18} +9\,{x}^{17} 
+7\,{x}^{16} +10\,{x}^{15} 
\nonumber \\ 
\hspace{-0.95in}&&   \quad \quad \, 
+7\,{x}^{13} +7\,{x}^{12}+8\,{x}^{10} +3\,{x}^{9} 
+8\,{x}^{8} +{x}^{6} +5\,{x}^{5} +2\,{x}^{4} +2\,{x}^{3} 
+{x}^{2} +2\,x +10, 
\nonumber 
\end{eqnarray}
\begin{eqnarray}
\label{v1mod11polbis}
\hspace{-0.95in}&&   \quad 
q_{9}\, \, \, = \,\,  \, \, 
2\,{x}^{45} +10\,{x}^{44} +8\,{x}^{43} +6\,{x}^{42} +8\,{x}^{41}
+8\,{x}^{40} +5\,{x}^{39} +{x}^{37} +5\,{x}^{36} +3\,{x}^{35}
\nonumber \\ 
\hspace{-0.95in}&&   \quad \quad\, 
+10\,{x}^{34}+2\,{x}^{33}
+3\,{x}^{32} +{x}^{31} +7\,{x}^{30} +8\,{x}^{28} +{x}^{27} 
+{x}^{25} +3\,{x}^{24} +5\,{x}^{22}
\nonumber \\ 
\hspace{-0.95in}&&   \quad \quad \, 
+9\,{x}^{21} +9\,{x}^{20} +4\,{x}^{19} +4\,{x}^{18} +2\,{x}^{17}
+8\,{x}^{16} +8\,{x}^{15} +2\,{x}^{14} +9\,{x}^{13} +9\,{x}^{12}
\nonumber \\ 
\hspace{-0.95in}&&   \quad \quad \, 
+6\,{x}^{11} +7\,{x}^{10} +3\,{x}^{9} +8\,{x}^{8} +{x}^{6}
+5\,{x}^{5} +2\,{x}^{4} +2\,{x}^{3} +{x}^{2} +2\,x +10,
\nonumber 
\end{eqnarray}
\begin{eqnarray}
\label{v1mod11polter}
\hspace{-0.95in}&&   \quad 
q_{8}\, \, \, = \,\,  \, \, 
6\,{x}^{45}+8\,{x}^{44}+2\,{x}^{43}+7\,{x}^{42}+2\,{x}^{41}
+4\,{x}^{40} +2\,{x}^{39}+7\,{x}^{37}+2\,{x}^{36}
\nonumber \\ 
\hspace{-0.95in}&&   \quad \quad\, 
+4\,{x}^{35} +2\,{x}^{34}+9\,{x}^{33}
+4\,{x}^{32}+2\,{x}^{31}+2\,{x}^{30}+2\,{x}^{29}+6\,{x}^{28}
+{x}^{27} +6\,{x}^{26}
\nonumber \\ 
\hspace{-0.95in}&&   \quad \quad \, 
+7\,{x}^{25}+9\,{x}^{24}+5\,{x}^{22}+10\,{x}^{21}+2\,{x}^{20}
+4\,{x}^{19}+5\,{x}^{18}+{x}^{17}+5\,{x}^{16}+3\,{x}^{15}
\nonumber \\ 
\hspace{-0.95in}&&   \quad \quad\, 
+7\,{x}^{14}
+3\,{x}^{13} +3\,{x}^{12} +10\,{x}^{11} +10\,{x}^{10} 
+3\,{x}^{9} +8\,{x}^{8} +{x}^{6} +5\,{x}^{5}
\nonumber \\ 
\hspace{-0.95in}&&   \quad \quad\, 
+2\,{x}^{4}+2\,{x}^{3}+{x}^{2}+2\,x+10,
\nonumber \\ 
\hspace{-0.95in}&&   \quad 
q_{7}\, \, \, = \,\,  \, \, 
2\,{x}^{40} +9\,{x}^{39} +4\,{x}^{37} +9\,{x}^{36} +6\,{x}^{35}
+5\,{x}^{34} +3\,{x}^{33} +6\,{x}^{32} +10\,{x}^{31}
\nonumber \\ 
\hspace{-0.95in}&&   \quad \quad\, 
+10\,{x}^{30} +10\,{x}^{29}+4\,{x}^{27}
+8\,{x}^{26}+4\,{x}^{25} +7\,{x}^{24} +7\,{x}^{23} +10\,{x}^{22}
+{x}^{21}
\nonumber \\ 
\hspace{-0.95in}&&   \quad \quad \, 
+8\,{x}^{20} +4\,{x}^{19} +10\,{x}^{18} +7\,{x}^{17} +{x}^{16} 
+10\,{x}^{14}  +6\,{x}^{13}  +6\,{x}^{12} +8\,{x}^{11} 
\nonumber \\ 
\hspace{-0.95in}&&   \quad \quad \, 
+3\,{x}^{10} +3\,{x}^{9}+8\,{x}^{8}
+{x}^{6} +5\,{x}^{5} +2\,{x}^{4} +2\,{x}^{3} +{x}^{2} +2\,x +10,
\nonumber 
\end{eqnarray}
\begin{eqnarray}
\label{v1mod11polter2}
\hspace{-0.95in}&&   \quad 
q_{6}\, \, \, = \,\,  \, \, (x+1)^{2} \cdot \, 
(10\,{x}^{38}+3\,{x}^{37}+6\,{x}^{36}+5\,{x}^{35}+7\,{x}^{34}
+3\,{x}^{33} +2\,{x}^{32}+5\,{x}^{31}
\nonumber \\ 
\hspace{-0.95in}&&   \quad \quad \, 
+10\,{x}^{30} +6\,{x}^{28} +9\,{x}^{27} +4\,{x}^{26}
+7\,{x}^{25} +{x}^{24}+2\,{x}^{23} +{x}^{22} +2\,{x}^{21}
+9\,{x}^{20} 
\nonumber \\ 
\hspace{-0.95in}&&   \quad \quad \, 
+8\,{x}^{19} +8\,{x}^{18} +2\,{x}^{17} +{x}^{16} +2\,{x}^{14} 
+3\,{x}^{13} +6\,{x}^{12} +6\,{x}^{11} +3\,{x}^{10} 
+8\,{x}^{9} 
\nonumber \\ 
\hspace{-0.95in}&&   \quad \quad \, 
 +4\,{x}^{8} +9\,{x}^{7} +8\,{x}^{6} +8\,{x}^{5}+10\,{x}^{4}
+10\,{x}^{3} +5\,{x}^{2} +4\,x+10), 
\nonumber \\ 
\hspace{-0.95in}&&   \quad 
q_{5}\, \, \, = \,\,  \, \, 
2\,{x}^{35} +8\,{x}^{34} +9\,{x}^{33} +2\,{x}^{32} +9\,{x}^{31}
+{x}^{30} +3\,{x}^{29} +2\,{x}^{28} +2\,{x}^{27} 
\nonumber \\ 
\hspace{-0.95in}&&   \quad \quad \, 
 +9\,{x}^{26}+{x}^{25}+7\,{x}^{24}
+6\,{x}^{23} +3\,{x}^{22} +8\,{x}^{21} +8\,{x}^{20} +4\,{x}^{19}
+4\,{x}^{18} +2\,{x}^{17}
\nonumber \\ 
\hspace{-0.95in}&&   \quad \quad \, 
+8\,{x}^{16} +8\,{x}^{15} +2\,{x}^{14} +9\,{x}^{13} +9\,{x}^{12}
+6\,{x}^{11} +7\,{x}^{10} +3\,{x}^{9}+8\,{x}^{8} +{x}^{6}
\nonumber \\ 
\hspace{-0.95in}&&   \quad \quad \, 
+5\,{x}^{5} +2\,{x}^{4}+2\,{x}^{3} +{x}^{2}+2\,x+10,
\nonumber 
\end{eqnarray}
\begin{eqnarray}
\label{v1mod11polter21}
\hspace{-0.95in}&&   \quad 
q_{4}\, \, \, = \,\,  \, \, (x+1) \cdot \,  
({x}^{34} +3\,{x}^{33} +7\,{x}^{32} +5\,{x}^{31} +5\,{x}^{30}
+8\,{x}^{29} +7\,{x}^{28} +{x}^{27} 
\nonumber \\ 
\hspace{-0.95in}&&   \quad \quad \, 
+6\,{x}^{26} +6\,{x}^{25}+8\,{x}^{24} +10\,{x}^{23} 
+4\,{x}^{22} +7\,{x}^{20} +{x}^{19}+3\,{x}^{18} +5\,{x}^{17}
\nonumber \\ 
\hspace{-0.95in}&&   \quad \quad \, 
+4\,{x}^{16}+3\,{x}^{15}+7\,{x}^{14}
+4\,{x}^{13}+3\,{x}^{12}+4\,{x}^{11}+7\,{x}^{10}+{x}^{9}
+2\,{x}^{8} +6\,{x}^{7}
\nonumber \\ 
\hspace{-0.95in}&&   \quad \quad \, 
+5\,{x}^{6}+7\,{x}^{5}+9\,{x}^{4} +4\,{x}^{3}+9\,{x}^{2} +3\,x+10),
\nonumber \\ 
\hspace{-0.95in}&&   \quad 
q_{3}\, \, \, = \,\,  \, \, 
5\,{x}^{30} +{x}^{29} +5\,{x}^{28} +9\,{x}^{27} +3\,{x}^{26}
+9\,{x}^{25} +10\,{x}^{24} +4\,{x}^{23} +6\,{x}^{21} 
\nonumber \\ 
\hspace{-0.95in}&&   \quad \quad \, 
+10\,{x}^{20} +4\,{x}^{19}+9\,{x}^{18}
+8\,{x}^{17}+4\,{x}^{16} +5\,{x}^{15} +5\,{x}^{14} +{x}^{13} 
+{x}^{12} 
\nonumber \\ 
\hspace{-0.95in}&&   \quad \quad \, 
+4\,{x}^{11} +3\,{x}^{9}
+8\,{x}^{8}+{x}^{6}+5\,{x}^{5}+2\,{x}^{4}
+2\,{x}^{3}+{x}^{2}+2\,x+10,
\nonumber 
 \end{eqnarray}
\begin{eqnarray}
\label{v1mod11polter22}
\hspace{-0.95in}&&   \quad 
q_{1}\, \, \, = \,\,  \, \, 
10\,{x}^{25} +8\,{x}^{24} +5\,{x}^{23} +7\,{x}^{22} +6\,{x}^{21}
+5\,{x}^{20} +4\,{x}^{19} +6\,{x}^{17} 
\nonumber \\ 
\hspace{-0.95in}&&   \quad \quad \, 
 +9\,{x}^{16} +6\,{x}^{15}
+4\,{x}^{14} +{x}^{11}
+6\,{x}^{10} +3\,{x}^{9} +8\,{x}^{8} +{x}^{6} +5\,{x}^{5}
\nonumber \\ 
\hspace{-0.95in}&&   \quad \quad \, 
+2\,{x}^{4} +2\,{x}^{3} +{x}^{2} +2\,x +10.
\nonumber 
\end{eqnarray}
More generally, one can conjecture that, 
modulo a prime $ p$, the corresponding polynomial 
equation will be of the form:
\begin{eqnarray}
\label{v1modp}
\hspace{-0.95in}&&     \, \, \, \, 
 \quad  \quad \quad  \quad \quad       \quad \quad      
\sum_{n=0}^{ p-1} \, q_{n} \cdot \,
  \Bigl(F(x)^{p-1} \Bigr)^n \,  \, \, \, = \,\,  \, \, \,  0.
\end{eqnarray}

\vskip .1cm

\section{Symmetries of the  differentially algebraic equation (\ref{H2mod5pol}).}
\label{compodiffalgapp}

Recall 
$   P(x,   F(x),  F'(x),  F''(x),  F'''(x),  F^{(4)}(x))
    =  \,   0$, i.e. the polynomial corresponding to 
equation (\ref{H2mod5pol}) for the differentially algebraic 
function (\ref{compoH1H2}) given by the
composition of two holonomic functions. It has been seen that
if $ F(x)$ is a solution of (\ref{H2mod5pol}), 
$ A \cdot  F(x)$ is also a solution of (\ref{H2mod5pol}). 
In fact this scaling symmetry is a consequence of a more general 
symmetry of the polynomial equation (\ref{H2mod5pol}).
If one considers the two solutions of the order-two
linear differential operator 
$ (16 x  -1)\cdot  x \cdot  D_x^2  +(32 x-1) \cdot D_x  +4$, 
namely $ S_1(x)  =   H_1(x)$ (see (\ref{H1H2}))
and $S_2(x)  =  \ln(x)\cdot  H_1(x)  +   {\cal H}_1(x)$
where the power series $ {\cal H}_1(x)$ reads:
\begin{eqnarray}
\label{qq}
\hspace{-0.98in}&&   
8\,x \, \,+84\,{x}^{2}\, \,\, +{\frac {2960}{3}} \,{x}^{3} \,\,\,
+{\frac {37310}{3}}\, {x}^{4} \,\,\,  \,+{\frac {820008}{5}} \,{x}^{5}
\, \,\,+{\frac {11153912}{5}} \,{x}^{6} \, \,
 \,+{\frac {1086209696}{35}}\, {x}^{7} 
\nonumber \\
\hspace{-0.98in}&&   \, \, \, 
 +{\frac {3074289075}{7}} \, {x}^{8}
 \,+{\frac {396822097100}{63}} \, {x}^{9} \,
+{\frac {28763739153292}{315}} \, {x}^{10} \,
+{\frac {73453759289456}{55}} \, {x}^{11}
\nonumber \\
\hspace{-0.98in}&&    \, \, \, 
+{\frac {9740489140942196}{495}}\, {x}^{12}\,
 \,+{\frac {375490923772997200}{1287}}\, {x}^{13}
\,\,\, \, + \, \, \, \cdots 
\end{eqnarray}
One finds that not only is $ S_1(H_2(x))$ a solution of the
non-linear ODE (\ref{H2mod5bis}), but $ S_2(H_2(x))$
is also a solution of the non-linear ODE (\ref{H2mod5bis}), 
and more generally, {\em any linear combination} 
 $  A \cdot  S_1(H_2(x))  +  B  \cdot   S_2(H_2(x)) $
is  a solution of  (\ref{H2mod5bis}). This is a general result: namely that for 
compositions of holonomic functions, the differentially
algebraic equation inherits the (differential Galois) symmetries 
of the linear differential operator corresponding to the first
 holonomic function.

\vskip .1cm

\section{Other solutions of the  differentially algebraic equation (\ref{H2mod5pol}).}
\label{Othersolapp}

\subsection{Movable singularities for the  differentially algebraic equation (\ref{H2mod5pol}).}
\label{Othersolapp1}

Among the miscellaneous solutions of (\ref{H2mod5bis}) one has a
one-parameter family such as:
\begin{eqnarray}
\label{oneparam}
\hspace{-0.95in}&&  \quad  
1 \,\, \,+4\,v \cdot \,x \,\, \,+4\,v \cdot \, (9\,v +4) \cdot \,{x}^{2} \,
\, \,+16\,v \cdot \, (25\,{v}^{2} +18\,v +9) \cdot \, {x}^{3}
\nonumber \\
\hspace{-0.95in}&&   \quad \quad \, 
 \,\, \,+4\,v \cdot \, (1225\,{v}^{3}+1200
\,{v}^{2}+792\,v+400) \cdot \, {x}^{4}
\nonumber \\
\hspace{-0.95in}&&   \quad \quad \, \, \,\,
+16\,v \cdot \, (3969\,{v}^{4}+4900\,{v}^{3}+3900\,{v}^{2}
  +2448\,v+1225) \cdot \, {x}^{5}
\nonumber \\
\hspace{-0.95in}&&   \quad\quad \,  \, \,\,
+ 16\,v \cdot \, (53361\,{v}^{5} +79380\,{v}^{4} +73500\,{v}^{3}
+53200\,{v}^{2}+32166\,v+15876)  \cdot \,  {x}^{6}
\nonumber \\
\hspace{-0.95in}&&   \quad \quad \quad  \quad \,\,  \,  \,\,
\,  \,+ \,\, \, \,\cdots 
\end{eqnarray}
which reduce to the series (\ref{seriescompoH1H2}) for 
$ v  =  1$. The radius of convergence, $ R$, of 
that series is clearly a function of the parameter 
$ v$ ($R  \rightarrow   0$ when 
$  v  \rightarrow   \infty$). 
One clearly has {\em movable singularities}. 

For $ v =  2  $ the series (\ref{oneparam}) reads:
\begin{eqnarray}
\label{v2}
\hspace{-0.95in}&&   \, 
1 \, \, \, \,+8\,x \,\,  \, \,+176\,{x}^{2}\, \, \,
+4640\,{x}^{3} \,\, \, +132672\,{x}^{4}\, \,
+3981600\,{x}^{5}\, \, \, +123476480\,{x}^{6}
\nonumber \\
\hspace{-0.95in}&&   \quad    \,+3921615488\,{x}^{7} \,
+126825870848\,{x}^{8} \,+4160174803232\,{x}^{9} \,
+138026770667008\,{x}^{10} 
\nonumber \\
\hspace{-0.95in}&&   \quad    \,+4622425971834496\,{x}^{11} \,
+156011453811509760\,{x}^{12}   \,
+5300311971681413248\,{x}^{13} 
\nonumber \\
\hspace{-0.95in}&&   \quad    \
\,+181089920155530497536\,{x}^{14}
 \,+ \,6217365460907222370816 \,{x}^{15} \, 
\,\, \, \, + \, \, \cdots 
\end{eqnarray}
This series has a radius of convergence 
$ R  \simeq  0.02722469  $
 corresponding to a singularity at $ x_c \simeq  0.02722469$.

\subsection{Reduction to algebraic functions modulo primes
for the other solutions of two differentially algebraic equation (\ref{H2mod5pol}).\\}
\label{reducmodapp}

\vskip .1cm

The $ v =  2$ series (\ref{v2}) 
is a solution of the  {\em non-linear} differential equation 
(\ref{H2mod5bis}), but {\em we have no idea if it can be seen 
as a composition of holonomic functions, 
and in particular a composition of 
diagonals of rational functions}.
Modulo $ 3$ this series (\ref{v2}) becomes
\begin{eqnarray}
\label{v2mod3}
\hspace{-0.95in}&&     \, \, \, \,  
1\,\,\,  +2\,x\,\,\,  +2\,{x}^{2}\, +2\,{x}^{3}\, +2\,{x}^{6}\, +2\,{x}^{7}\, 
+2\,{x}^{8}\, +2\,{x}^{9}\, +{x}^{10}\, +{x}^{11}\, +{x}^{13}\, +{x}^{14}
\nonumber \\
\hspace{-0.95in}&&   \quad    \quad  \, \,  \, 
\, +2\,{x}^{16}
\, +2\,{x}^{17}\, +2\,{x}^{18}\, +2\,{x}^{21} \, 
+2\,{x}^{24} \, +2\,{x}^{25} \, +2\,{x}^{26}\, +2\,{x}^{27} 
\, \, \,+ \,\,  \, \cdots 
\end{eqnarray}
which can be seen to be
 an algebraic solution, modulo $ 3$, of 
\begin{eqnarray}
\label{H2mod3bis}
\hspace{-0.95in}&&   \quad \quad \,  
({x}^{2}+2\,x+2)  \cdot \, F(x)^{4} \, \,  
\, +2 \cdot \, (x+1) \cdot \, F(x)^2 \,  \, \, +2 \cdot \, (x+1)
\,  \,\,   = \,\, \,   \, \,  0,
\end{eqnarray}
which is  the same polynomial
 equation as (\ref{H2mod3}) (even if the series (\ref{v1mod3}) 
and (\ref{v2mod3}) are different). 
Modulo $ 5$  the series (\ref{v2}) becomes
\begin{eqnarray}
\label{v2mod5}
\hspace{-0.95in}&&     \, 
1 \, \,  +3\,x \, \,  +{x}^{2} \, +2\,{x}^{4}+3\,{x}^{7}+3\,{x}^{8}
+2\,{x}^{9}+3\,{x}^{10}+{x}^{11}+3\,{x}^{13}+{x}^{14}\, 
+{x}^{15}\, +3\,{x}^{16}
\nonumber \\
\hspace{-0.95in}&&   \quad    \quad  \, \,  \, \, \, 
+4\,{x}^{17}\, +{x}^{18}\, +2\,{x}^{20}\, +4\,{x}^{21}\, +{x}^{22}\, 
+4\,{x}^{23}\, +{x}^{26}\, +4\,{x}^{27}\, \,\, + \,\,  \, \cdots 
\end{eqnarray}
which similarly satisfies, modulo $ 5$ the algebraic equation
\begin{eqnarray}
\label{H2mod5bisbis}
\hspace{-0.95in}&&   \quad \quad \quad 
({x}^{8}+{x}^{6}+4\,{x}^{5}+2\,{x}^{4}+3\,{x}^{3}+3\,{x}^{2}+3\,x+1)
 \cdot \, F(x)^{16} 
\nonumber \\ 
\hspace{-0.95in}&&   \quad  \quad \quad \quad \quad 
\, + \, ({x}^{4}+3\,{x}^{3}+3\,{x}^{2}+3\,x+1) 
 \cdot \,  (F(x)^{12}\, +F(x)^{4}\, +1)
  \\ 
\hspace{-0.95in}&&   \quad  \quad \quad \quad \quad \,
+ \, ({x}^{2}-x+1)  \, (3\,{x}^{4}+5\,{x}^{3}+6\,{x}^{2}+4\,x+1)
 \cdot \,  F(x)^{8}
 \,\,  \, \, = \, \,  \,\,  0,
  \nonumber
\end{eqnarray}
which is actually the same polynomial
 equation as (\ref{H2mod5}) (even if the series 
(\ref{seriescompoH1H2mod5}) and (\ref{v2mod5}) are different). 
Modulo $ 7$  the series (\ref{v2}) becomes
\begin{eqnarray}
\label{v2mod7}
\hspace{-0.95in}&&  \quad   1 \, +x \, +{x}^{2} \, +6\,{x}^{3} \, +{x}^{4}
 \, +{x}^{6} \, +5\,{x}^{8}  \, +6\,{x}^{10} \, +2\,{x}^{12} \, +2\,{x}^{13}
 \, +5\,{x}^{14} \, +4\,{x}^{15}
\nonumber \\
\hspace{-0.95in}&&   \quad   \quad     \, \,  \, \, \, 
+6\,{x}^{16} \, +2\,{x}^{17} \, 
+2\,{x}^{17} \,  +6\,{x}^{18} \, +3\,{x}^{19} 
\, +6\,{x}^{20} \, +3\,{x}^{21} \, +4\,{x}^{22}
\nonumber \\ \quad  
\hspace{-0.95in}&&   \quad   \quad     \, \,  \, \, \, 
+6\,{x}^{23} \, +5\,{x}^{24} \, +3\,{x}^{25}
 \,\, + \, \, \cdots 
\end{eqnarray}
which similarly satisfies, modulo $7 $ the 
same algebraic equation as (\ref{v1mod7}).
Modulo $ 11$  the series (\ref{v2}) can actually be seen to be
 an algebraic function, modulo $ 11$, of the 
{\em same polynomial equation as} (\ref{v1mod11}).

One can expect the series (\ref{v2}) 
{\em to reduce to algebraic functions modulo every prime or
power of a prime},  but not necessarily with the same polynomial 
equations when the primes become large enough.
 
\vskip .2cm

{\bf Remark:} Actually the series (\ref{oneparam})
seems to reduce to algebraic functions modulo every prime or
power of a prime for many prime-integer values of $ v$, even  many
rational values of $ v$, with the polynomial equations
also reducing to the previous polynomials  (\ref{H2mod5}), 
(\ref{H2mod3}),  (\ref{v1mod7}),  (\ref{v1mod11}). 

Such simple results provide a strong incentive to  
systematically study the reduction, modulo primes or powers of a prime,  
of all solution-series with integer coefficients of differentially 
algebraic equations. 

\vskip .1cm

\section{ Linear diff-Pad\'e approach of a differentially algebraic function:
fixed singularities versus movable singularities.}
\label{transcendantdiffPade}

\vskip .1cm

The traditional  diff-Pad\'e approach amounts to seeking linear 
differential equations with {\em polynomial coefficients}. However, 
from a more formal (and less well-defined) viewpoint, if one seeks 
 linear differential equations with no longer polynomial functions
but holonomic function coefficients, it is easy to see that 
the (differentially algebraic) function (\ref{compoH1H2}) 
can also be seen as a  solution of the following linear 
second order differential 
operator but with {\em transcendental} coefficients
\begin{eqnarray}
\label{translin}
\hspace{-0.95in}&&   \,    \quad  \quad  
{{d p(x)} \over {dx}} \cdot \, p(x)
 \cdot \, (1 \, -16\,p(x)) \cdot \, D_x^2 
\nonumber \\
 \hspace{-0.95in}&&  \quad   \quad \quad  \quad  \quad \,  \, 
\, -\Bigl( p(x) \cdot \, 
(1- \, 16 \, p(x)) \cdot \, {{d^2 p(x)} \over {dx^2}} \,
 - (1 \, -32 \,p(x)) \cdot \,  \Bigl({{d p(x)} \over {dx}}\Bigr)^2
\Bigr) \cdot \, D_x
\nonumber \\
 \hspace{-0.95in}&&  \quad    \quad \quad  \quad   \, \, 
 \quad  \quad  \, -4 \cdot \, \Bigl({{d p(x)} \over {dx}}\Bigr)^3 ,
\end{eqnarray}
where the pullback $ p(x)$ is the transcendental function:
\begin{eqnarray}
\label{pullp}
\hspace{-0.95in}&&  \quad  \quad  \quad  \quad  \quad  \quad  \quad  
p(x)\, \, = \, \, \,  x \cdot \,
 _2F_1\Bigl(\Bigl[{{1} \over {2}}, {{1} \over {2}}\Bigr], \, [1], \, 16 \, x\Bigr).
\end{eqnarray}
It is quite easy to verify that (\ref{seriescompoH1H2}), 
the series expansion of the differentially algebraic function (\ref{compoH1H2}), 
is actually a solution of the formal transcendental 
linear differential equation (\ref{translin}).  
The singularities of 
(\ref{translin}) are fixed, and correspond to the 
head coefficient, namely:
\begin{eqnarray}
\label{headtrans}
\hspace{-0.95in}&&   \quad  \quad  \quad  \quad  \quad  \quad  \quad  
{{d p(x)} \over {dx}} \cdot \, p(x) \cdot \, (1 \, -16\,p(x)) 
\, \, \, = \, \, \, \, 0.
\end{eqnarray}
At first sight one expects an infinite number of zeros from
the vanishing condition (\ref{headtrans}) of {\em transcendental} 
functions. Here, it seems that the only solutions of (\ref{headtrans}) 
(fixed singular points of (\ref{translin})) are $ 0$ and the 
previously given value $ x    \simeq   0.04602833718455$. 

\vskip .1cm

Keeping this very formal result in mind, let us display the kind of 
result we obtained for a ``traditional'' linear diff-Pad\'e 
approximant analysis (the coefficients are now polynomials).
Recall that near $\, x\,  \simeq \, 1/16$, 
$\, 16 \, x \cdot \, _2F_1([1/2, 1/2], \, [1], \, 16 \, x)$ behaves like 
$ \, p(x) \, \simeq \, -\, 16 \, x \cdot \, \ln(1\, -16\, x)/2$, and  
the (differentially algebraic) function (\ref{compoH1H2})
behaves like $\, _2F_1([1/2, 1/2], \, [1], \, \, p(x))$
with $\, p(x) \, \rightarrow \, \infty$, namely like  
$\, \ln(-p(x))/2/(-p(x))^{1/2}$, which is an irregular singularity.

The order four and eight differential approximants give a singularity 
extremely close to the previous exact value with an exponent 
very close to zero ($.005353572$). 

One also finds many singularities very close to $\, 1/16$, 
namely  $\, .062500000019$ 
$ \pm \, 1.327 \, 10^{-11}$
$\, .06249973$,  $\, .0624999976$,  $\, .062499999808$, ...
the exponents being all over the place and changing from 
approximant to approximant. This type of behaviour  of differential 
approximants  is characteristic of an irregular singularity.

\vskip .1cm

\vskip .5cm
\vskip .5cm


\end{document}